\documentclass[twocolumn,superscriptaddress,amsfont,amssymb,amsmath, showpacs,balancelastpage, nofootinbib]{revtex4-1}

\usepackage{CJKutf8}

\usepackage{amsthm}

\usepackage[utf8]{inputenc}
\usepackage{amsmath,amssymb,amsfonts,stmaryrd}
\usepackage{physics}
\usepackage{dsfont} 
\usepackage{graphicx}
\usepackage{xcolor}
\usepackage{graphicx}
\usepackage{cancel}
\usepackage{natbib}
\usepackage{color}
\usepackage{tikz}
\usepackage{mathrsfs}
\usepackage{relsize}
\usepackage{multirow}
\usepackage[linktocpage]{hyperref}
\usepackage[all]{xy}
\hypersetup{colorlinks=true,citecolor=blue,linkcolor=blue, urlcolor=blue, breaklinks=true}

\usepackage[mathscr]{euscript}
\usepackage[scr=boondox]{mathalpha}
\usepackage{comment}

\usepackage{float}

\usepackage{bm} %allows for bold math type
\usepackage{subfigure} %allows for 2x2 figures
\usepackage{environ}
\usepackage{url}
\usepackage[margin=0.75in]{geometry}

\usepackage{mathtools}

\newcommand{\Z}{\mathbb{Z}}
\newcommand{\OO}{\text{o}}

\newcommand{\SO}{\mathscr{S}_{\text{o}}}
\newcommand{\PO}{\vec{\mathscr{P}}_{\text{o}}}

\begin{document}

\title{
Electric polarization and discrete shift
from boundary \\ and corner charge in crystalline Chern insulators 
}

\author{Yuxuan Zhang}
\author{Maissam Barkeshli}
\affiliation{Department of Physics and Joint Quantum Institute University of Maryland,
College Park, Maryland 20742, USA}

\begin{abstract}
Recently, it has been shown how topological phases of matter with crystalline symmetry and $U(1)$ charge conservation can be partially characterized by a set of many-body invariants, the discrete shift $\mathscr{S}_{\OO}$ and electric polarization $\vec{\mathscr{P}}_{\OO}$, where $\OO$ labels a high symmetry point. Crucially, these can be defined even with non-zero Chern number and/or magnetic field.  One manifestation of these invariants is through quantized fractional contributions to the charge in the vicinity of a lattice disclination or dislocation. In this paper, we show that these invariants can also be extracted from the length and corner dependence of the total charge (mod 1) on the boundary of the system. We provide a general formula in terms of $\mathscr{S}_{\OO}$ and $\vec{\mathscr{P}}_{\OO}$ for the total charge of any subregion of the system which can include full boundaries or bulk lattice defects, unifying boundary, corner, disclination, and dislocation charge responses into a single general theory. These results hold for Chern insulators, despite their gapless chiral edge modes, and for which an unambiguous definition of an \it intrinsically two-dimensional \rm electric polarization has been unclear until recently.
We also discuss how our theory can fully characterize the topological response of quadrupole insulators. 

\end{abstract}

\maketitle

\tableofcontents

\section{Introduction}

Over the past few decades, substantial progress has been made in the understanding of (2+1)D topological phases of matter with crystalline symmetry (for a partial list of references, see for example \cite{wen2002quantum,wen04,hasan2010,fu2011topological,barkeshli2012a,Essin2013SF,Essin2014spect,barkeshli2019,Benalcazar2014,ando2015,YangPRL2015,watanabe2015filling,watanabe2016filling,Chiu2016review,hermele2016,barkeshli2019tr,zaletel2017,Po2017symmind,song2017,Huang2017,Shiozaki2017point,Kruthoff2017TCI,Bradlyn2017tqc,schindler2018higher,watanabe2018,Miert2018dislocationCharge,khalaf2018symmetry,Thorngren2018,tang2019comprehensive,Liu2019ShiftIns,Song2020,Li2020disc,manjunath2021cgt,manjunath2020FQH,Cano_2021,Elcoro2021tqc,manjunath2022mzm,herzogarbeitman2022interacting,zhang2022fractional,zhang2023complete,manjunath2023classif,sachdev2023quantum,kobayashi2024,zhang2023complete,zhang2022pol,manjunath2024Characterization,kobayashi2024crystalline}). In particular, recently a number of topological invariants protected by crystalline symmetry have been understood which are well-defined in the many-body interacting setting beyond single-particle band theory, and which correspond to quantized physical responses \cite{manjunath2021cgt,manjunath2020FQH,zhang2022fractional,zhang2022pol,zhang2023complete}. In this paper, we focus on two such invariants, which we refer to as the discrete shift $\mathscr{S}_{\OO}$ and the electric (charge) polarization $\vec{\mathscr{P}}_{\OO}$, where $\OO$ denotes a high symmetry point in the unit cell. Ref. \cite{zhang2022fractional,zhang2022pol} showed how to extract these many-body invariants from microscopic models and precisely match predictions from topological quantum field theory and G-crossed braided tensor category theory \cite{manjunath2020FQH,manjunath2021cgt,barkeshli2019}. Crucially, these results apply also in the case of non-zero Chern number and/or magnetic field. 

The discrete shift $\mathscr{S}_{\OO}$ is a $\mathbb{Z}_M$ invariant protected by $M$-fold rotations about $\OO$, and it specifies a quantized fractional contribution to the electric charge in the vicinity of a lattice disclination centered at $\OO$.\cite{zhang2022fractional,zhang2022pol} It also specifies a dual response, the angular momentum of magnetic flux, and can be extracted from (partial) rotation operations.\cite{zhang2023complete} 

The electric polarization $\vec{\mathscr{P}}_{\OO}$ can be viewed as a topological invariant associated with translational symmetry. It is quantized in the presence of $M$-fold rotational symmetry, and can only take non-trivial quantized values when $M = 2,3,4$ \cite{manjunath2021cgt}. In the absence of rotational symmetry ($M = 1$), $\vec{\mathscr{P}}_{\OO}$ can be viewed as an unquantized topological response \cite{Song2021polarization}. We emphasize that $\vec{\mathscr{P}}_{\OO}$ is an \it intrinsically two-dimensional \rm polarization, not an effective 1d polarization of the 2d system viewed as a 1d system, as is often considered in discussions of Chern insulators. 

The electric polarization is of particular interest, because the question of whether electric polarization can be defined in Chern insulators has been somewhat unclear until recently. Ref. \cite{coh2009} provided a single-particle Berry phase definition of electric polarization in Chern insulators, but this requires an arbitrary choice of momentum in the Brillouin zone, whose physical meaning is unclear. Ref. \cite{Fang2012PGS,Song2021polarization} later suggested that electric polarization may not be well-defined in Chern insulators \footnote{See e.g. Table I of \cite{Fang2012PGS} and  first paragraph of Appendix C of 
\cite{Song2021polarization}}. Recently \cite{zhang2022pol} showed unambiguously that one can define an electric polarization in Chern insulators consistently through a variety of different physical response properties of the system. For this paper, the most relevant of these is that it specifies a fractional quantized contribution to the charge in the vicinity of a lattice defect with non-zero Burgers vector, such as a lattice dislocation or an impure lattice disclination. 

In the case of Chern number $C = 0$, it is known that the discrete shift and electric polarization have implications for the boundary and corner charge of the system. In particular, the fractional charge associated with a lattice disclination also implies fractional charge at corners of the system \cite{Benalcazar2019HOTI,Li2020disc,naren2023rotational,rao2023effective,maymann2022prb}. Similarly, electric polarization is well-known to specify the boundary charge density. 

The purpose of this paper is to study the fate of these corner and boundary charges in the case of non-zero Chern number, $C \neq 0$, where the system has topologically protected gapless edge states. Specifically, to what extent can $\mathscr{S}_{\OO}$ and $\vec{\mathscr{P}}_{\OO}$ be extracted from the boundary and corner charge of Chern insulators with crystalline symmetry? 

The main result of this paper is Eq. \ref{eq:wt}-\ref{eq:charge}, which gives the total charge (mod 1) on the boundary of a Chern insulator with crystalline symmetry in terms of quantized topological invariants, and which is invariant to any local perturbations on the boundary. We derive Eq. \ref{eq:charge} from topological quantum field theory considerations and match it to numerical calculations on microscopic models. $\mathscr{S}_\OO$ contributes to the corner-angle dependence of the total charge mod 1 while $\vec{\mathscr{P}}_{\OO}$ contributes to the length-dependence along the boundary. The choice of high symmetry point $\OO$ manifests as a specific ambiguity in decomposing various  contributions to the boundary charge. These results unify the boundary, corner, disclination, and dislocation charge responses into a single general theory. 

Our results suggest that  $\vec{\mathscr{P}}_{\OO}$ may be experimentally measurable in crystalline Chern insulators from high-resolution scanning local charge measurements along the boundary of two-dimensional quantum materials. Our results also suggest a variety of other geometries that could be used to infer the corner-angle dependence of the boundary charge and extract $\mathscr{S}_{\OO}$.

One application of our general theory is in giving a complete characterization of quadrupole insulators and related higher-order topological insulators (HOTIs). In particular, the quantized corner charge is extensively studied in the HOTI literature \cite{wladimir2017quantized,schindler2018higher,Benalcazar2019HOTI,roy2020dislocation,maymann2022prb,hirsbrunner2023crystalline,khalaf2018higher}, and has been explained using multipolar moment. Recently, it has been shown that multipolar moment is inadequate to account for the corner charges \cite{Jahin2024HOTI}. We show that the corner charge response can be fully accounted for by the discrete shift $\SO$.

\subsection{Organization of paper}

The remainder of this paper is organized as follows. Sec.~\ref{sec:main_result} defines the charge response to boundaries and bulk defects, which is the main result of our paper. 
Sec.~\ref{sec:response_quantities} defines the relevant geometrical measures and the notion of extra flux $\delta\Phi_{W,\OO}$ of the boundary and bulk defects. Sec.~\ref{sec:charge}  presents the numerical calculations for the square lattice Hofstadter model that verify our main result. Sec.~\ref{sec:edgeC} outlines the procedure for calculating $\PO$ through edge charge on one boundary of a cylinder. Sec.~\ref{sec:cornerC} presents details on calculating $\SO$ through corner contributions to the charge. Sec. \ref{sec:general} establishes an equivalence between corners and disclinations; edges and dislocations. Sec.~\ref{sec:field} reviews the derivation of the charge response using the framework of topological quantum field theory. Sec.~\ref{sec:HOTI} applies our charge response to a HOTI model and calculates its $\SO$ and $\PO$.

\begin{figure}[t]
    \centering
    \includegraphics[width=2.5cm]{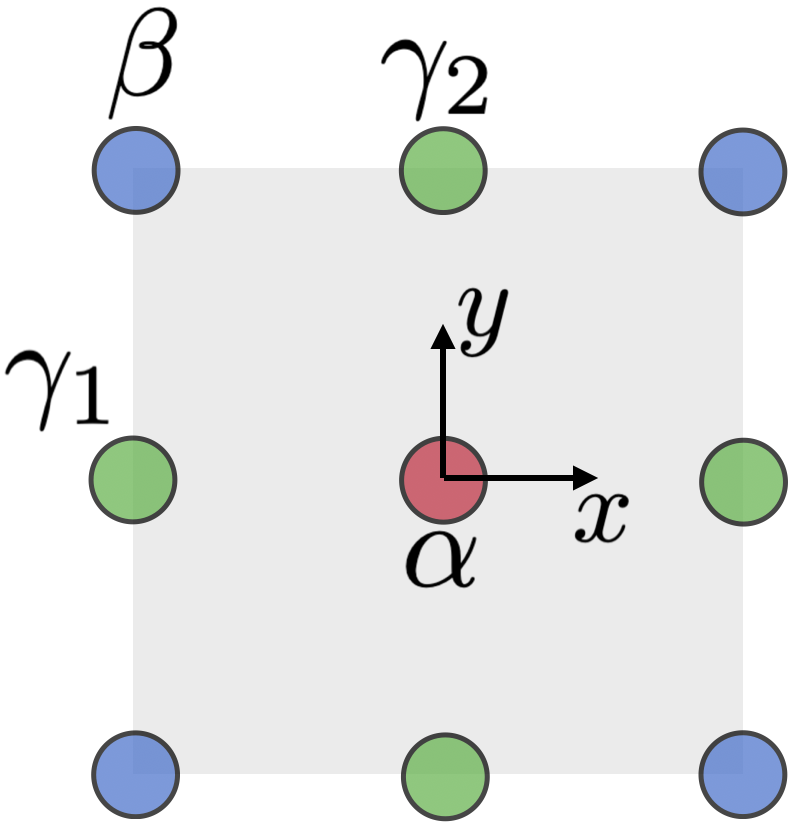}
    \caption{The $C_4$ symmetric unit cell with maximal Wyckoff positions $[\OO]\in\{\alpha, \beta, \gamma_1, \gamma_2\}$. The choice of unit cell in the square lattice is in general arbitrary, and we use the convention that $\beta$ represents vertices and $\alpha$ represents plaquette centers.}
    \label{fig:unit_cell}
\end{figure}

\begin{figure*}[t]
    \centering
    \includegraphics[width=17.5cm]{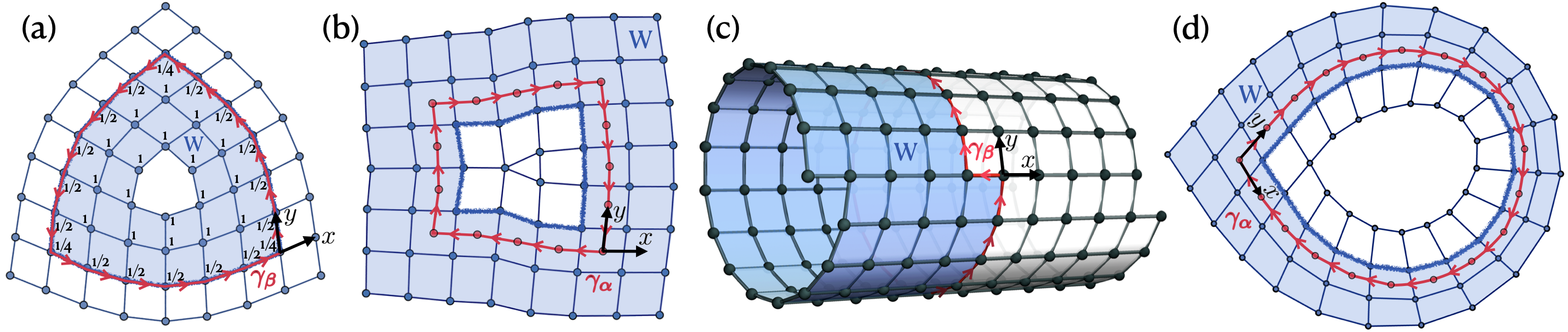}
    \caption{\textbf{(a)} Region $W$ covering a boundary with three corners. The red loop $\gamma_{\beta}$ is aligned with $\partial W$ and determines the corresponding $\Gamma=\pi/2$ and $\vec{L}_{\beta}=(-6,0)$, the total corner angle is $\Omega_{\text{cor}}=\Gamma-2\pi=-3\pi/2$. The weightings for $Q_W$ are labeled on each site. This lattice is created by removing the central site of a $\Omega_{\text{disc}} = \frac{\pi}{2}$ disclination. \textbf{(b)} Region $W$ covering a boundary with four corners. The red loop $\gamma_{\alpha}$ determines the corresponding $\Gamma=4\pi$ and $\vec{L}_{\OO}=(0,-1)$, the total corner angle is $\Omega_{\text{cor}}=\Gamma-2\pi=2\pi$. \textbf{(c)} Region $W$ covering one side of a cylinder with non-trivial shear. Here, $\Gamma = 2\pi$ and  $\vec{L}_{\beta}=(16,-1)$ \textbf{(d)} Region $W$ covering the outside of a ribbon. Here, $\Gamma=5\pi/2$, $\vec{L}_{\alpha}= (0,19)$, and the corner angle is $\Omega_{\text{cor}}=\pi/2$. }
    \label{fig:lattices}
\end{figure*}

\section{Main Result}
\label{sec:main_result}

We consider a Chern insulator with $U(1)$ charge conservation symmetry, $\mathbb{Z}^2$ (magnetic) translation symmetry with flux $\phi$ per unit cell, and a $\mathbb{Z}_4$ rotational symmetry. The full symmetry group we consider is then $G = U(1) \times_\phi [\mathbb{Z}^2 \rtimes \mathbb{Z}_4]$. The Chern insulator has a Chern number $C$ and a charge per unit cell (filling) $\nu$. 

Refs. \cite{zhang2022fractional, zhang2022pol} showed the existence of quantized topological invariants $\mathscr{S}_{\OO}$ and $\vec{\mathscr{P}}_{\OO}$\footnote{We caution that we have slightly abused terminology, as our $\PO$ is related to the conventional definition of polarization $\vec{P}_{\OO}$ by 
$\PO=(\mathscr{P}_{\OO,x},\mathscr{P}_{\OO,y})=(P_{\OO,y},-P_{\OO,x})=\vec{P}_{\OO}\times \hat{z}$, as explained in \cite{zhang2022pol}.}, which depend on a maximal Wyckoff position (MWP) $[\OO] = \alpha, \beta,\gamma_1,\gamma_2$ (see Fig. \ref{fig:unit_cell}). For a fixed $C$, $\mathscr{S}_{\OO}$ can take four possible values modulo $4$, such that $\mathscr{S}_{\OO} \mod 1 = \frac{C}{2} \mod 1$. The quantized electric polarization $\vec{\mathscr{P}}_{\OO}$ defines a $\mathbb{Z}_2$ invariant. For the $C_4$ symmetric MWPs $\alpha, \beta$, we have $\vec{\mathscr{P}}_{\OO} = (0,0)$ or $(1/2, 1/2) \mod \mathbb{Z}^2$. The dependence on $\OO$ was found to be \cite{zhang2022pol}:
\begin{equation}\label{eq:S_P_relations}
    \{\mathscr{S}_{\beta},\vec{\mathscr{P}}_{\beta},\kappa\}=\{\mathscr{S}_{\alpha}+4\mathscr{P}_{\alpha,y}-\kappa,\vec{\mathscr{P}}_{\alpha}+(\frac{\kappa}{2},\frac{\kappa}{2}),\kappa\} ,
\end{equation}
where $\kappa \equiv \nu - C \phi/2\pi$. Note that the fact that electric polarization of a system with total non-zero charge requires a choice of origin $\OO$ in the unit cell is well-known. The dependence on $\OO$ is usually removed when a neutralizing background, such as a background ionic contribution, is added. In this paper we are focused on properties of the electronic system, and thus do not consider a neutralizing background.

To determine the contribution of these invariants to the charge response, we consider a large subregion $W$ of the system. $W$ is chosen such that its boundary $\partial W$ is deep in the bulk, far away from any boundaries of the lattice and any defects in the interior of the lattice. Moreover, $W$ is defined so that $\partial W$ is aligned with the boundary of the unit cell. We note that unlike the definition in Ref. \cite{zhang2022pol}, $W$ can include boundaries and corners of the lattice in addition to disclinations and dislocations. Our results show how equivalences can be made between lattice defects and boundaries, which will be discussed in Sec.~\ref{sec:general}.

The total charge $Q_W$ within the region $W$ is defined as \cite{zhang2022fractional}
\begin{equation}\label{eq:wt}
    Q_{W}\equiv\sum_{i\in W} \text{wt}(i)Q_i. 
\end{equation}
Here $i \in W$ labels the sites in $W$, and $Q_i$ is the average charge on site $i$. The weighting factor $\text{wt}(i) = 1$ if $i$ 
is in the interior $W$, and $\text{wt}(i)=0$ if $i$ is outside of $W$. For sites $i$ that lie at the boundary $\partial W$, $2\pi\text{wt}(i)$ is the angle subtended by $\partial W$ in the interior of $W$ at $i$. 

We find that, in the limit where $\partial W$ is far from boundaries and defects, $Q_{W}$ obeys the following equation:
\begin{equation}\label{eq:charge}
    Q_{W} = 
    \SO\frac{\Gamma}{2\pi}+\vec{L}_{\OO}\cdot\vec{\mathscr{P}}_{\OO} + \nu n_{W,\OO}+\frac{C\delta\Phi_{W,\OO}}{2\pi}\mod 1.
\end{equation}
Here, the quantities $\Gamma$, $\vec{L}_{\OO}$, $n_{W,\OO}$, and $\delta \Phi_{W,\OO}$ depend on geometrical properties of the lattice with boundaries, corners, dislocations, and disclinations in the region $W$, and will be defined precisely in Sec.~\ref{sec:response_quantities}. Briefly, $2\pi - \Gamma$ is the total angle by which a vector is rotated upon traversing $\partial W$.\footnote{As we will explain, $\Gamma$ is defined as a real number, not just modulo $2\pi$. } $\vec{L}_{\OO}$ is the sum of translation vectors obtained upon traversing a loop $\gamma_{\OO}$ in $W$ that starts at $\OO$ and encloses all lattice boundaries and defects in $W$.
$\gamma_{\OO}$ should be smoothly deformable to the boundary $\partial W$ without passing through any defects or boundaries of the lattice. For simplicity we also require $\gamma_{\OO}$ to be non-self-intersecting. $n_{W,\OO}$ is a measure of an effective number of unit cells in $W$, and $\delta \Phi_{W,\OO}$ is a measure of the change in magnetic flux in $W$ relative to an appropriate reference background.

The main point of Eq. \ref{eq:charge} is that given a choice of high symmetry point $\OO$, there are distinct fractionally quantized contributions to $Q_W$ arising from the invariants $\mathscr{S}_{\OO}$ and $\vec{\mathscr{P}}_{\OO}$, Chern number $C$, and filling $\nu$.  

Eq.~\eqref{eq:charge} is numerically verified in Sec.~\ref{sec:edgeC}, \ref{sec:cornerC}, and \ref{sec:HOTI}, and is explained using field theory in Sec.~\ref{sec:field}.

\section{Geometrical Measures}\label{sec:response_quantities}

\begin{figure*}[t]
    \centering
    \includegraphics[width=12cm]{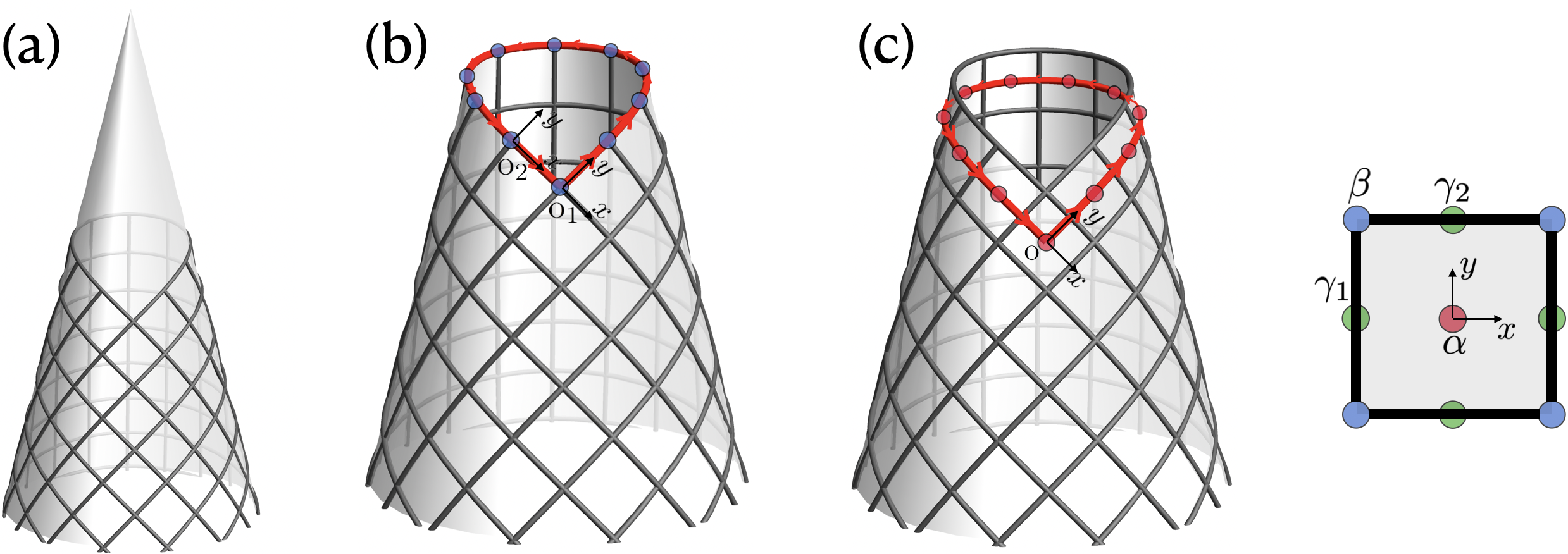}
    \caption{\textbf{(a)} Lattice with corner angle $\Gamma=-\pi/2$ can be isometrically embedded on a cone with apex angle $\Omega=3\pi/2$ (i.e. disclination angle for a lattice).  \textbf{(b)} $\Gamma$ and $\vec{L}_{\OO}$ can be calculated using the red loop via Eq.~\ref{eq:lo}. For $\OO=\beta$, choosing origin to be either $\OO_1$ or $\OO_2$ gives either $\vec{L}_{\OO_1}=(0,10)$ or $\vec{L}_{\OO_2}=(1,9)$, which is in the same equivalence class. \textbf{(c)} For $\OO=\alpha$, $\vec{L}_{\OO}=(0,11)$ which lies in a different equivalence class compared to choosing $\OO=\beta$.}
    \label{fig:cones}
\end{figure*}

In this section we define precisely the geometrical quantities $\vec{L}_{\OO}$, $\Gamma$, $n_{W,\OO}$ and $\delta \Phi_{W,\OO}$ used in Eq. \ref{eq:charge}, and their relationship to Burger's vectors and Frank angles of lattice dislocations and disclinations.

\subsection{Definitions of $\vec{L}_{\OO}$ and $\vec{b}_{\OO}$}

Consider a loop $\gamma_{\OO}$ which is obtained by starting and ending at a high symmetry point $\OO$ and following a set of unit translation vectors. 

Note that $\OO$ refers to a point on the lattice. We can write $[\OO] \in \{\alpha, \beta, \gamma_1, \gamma_2\}$ as the maximal Wyckoff position (MWP) of $\OO$. In this paper we will slightly abuse notation and drop the square brackets. Whether a given quantity depends on $\OO$ as a specific high symmetry point in the lattice or only through its MWP should be clear from context. 

The interior of $\gamma_{\OO}$ contains all relevant defects and boundaries whose charge response we wish to compute using Eq. \ref{eq:charge}. Note that here the interior is  defined to the left of the loop; that is, in the direction of the cross product of the out-of-plane direction and the translation vector. 
We then define
\begin{align}\label{eq:lo}
   \vec{L}_{\OO}&\equiv \sum_{j \in \gamma_{\OO}}\hat{L}_j.
\end{align}
Here the sum is taken over the set of unit translations needed to traverse $\gamma_{\OO}$,
with $\hat{L}_j\in \{\pm\hat{x},\pm\hat{y}\}$ being the unit translation vectors, and $j$ being points on the loop $\gamma_{\OO}$ related by the translations. 
All $j$ points correspond to the same MWP as $\OO$.

When $\gamma_{\OO}$ encloses a single boundary, $\vec{L}_{\OO}$ is a vectorized edge length of that boundary. In Fig.~\ref{fig:lattices}(c) we give an example of calculating $\vec{L}_{\OO}$ on a cylinder with non-zero shear. In this case $\vec{L}_{\OO}$ is independent of $\OO$, and the subscript can be omitted. When $\gamma_{\OO}$ encloses a single dislocation or disclination, $\vec{L}_{\OO}$ is reduced to the Burgers vector $\vec{b}_{\OO} = \vec{L}_{\OO}$ (see also the discussion below in Sec. \ref{sec:equivalence}).

Recall that, as discussed in \cite{zhang2022pol}, the Burgers vector for a pure dislocation, in the absence of any disclinations, is independent of the choice of origin $\OO$. However in the presence of disclinations, the Burgers vector does depend on the MWP of $\OO$.

\subsection{Definitions of $\Gamma$, $\Omega_{\text{disc}}$ and $\Omega_{\text{cor}}$}

For the loop $\gamma_{\OO}$, $\Gamma$ is defined as: 

\begin{align}\label{eq:gamma}
   \Gamma \equiv 2\pi - \sum_{j \in \gamma_{\OO}} K_j.
\end{align}
Here, the $\sum_{j \in \gamma_{\OO}}$ is as above, where we sum over points related by unit translation vectors. $K_j$ is the curvature of the loop at the point $j$ on $\gamma_{\OO}$. More specifically, $K_j$ is equal to $\pi-\theta$ where $\theta \in \{\pi/2,\pi, 3\pi/2\}$ is the angle subtended by the inside of the loop. Importantly, we define $\theta$ as a real number (not just modulo $2\pi$), so we have chosen a particular lift of the angles to the real numbers.  

\begin{figure}[t]
    \centering
    \includegraphics[width=5.5cm]{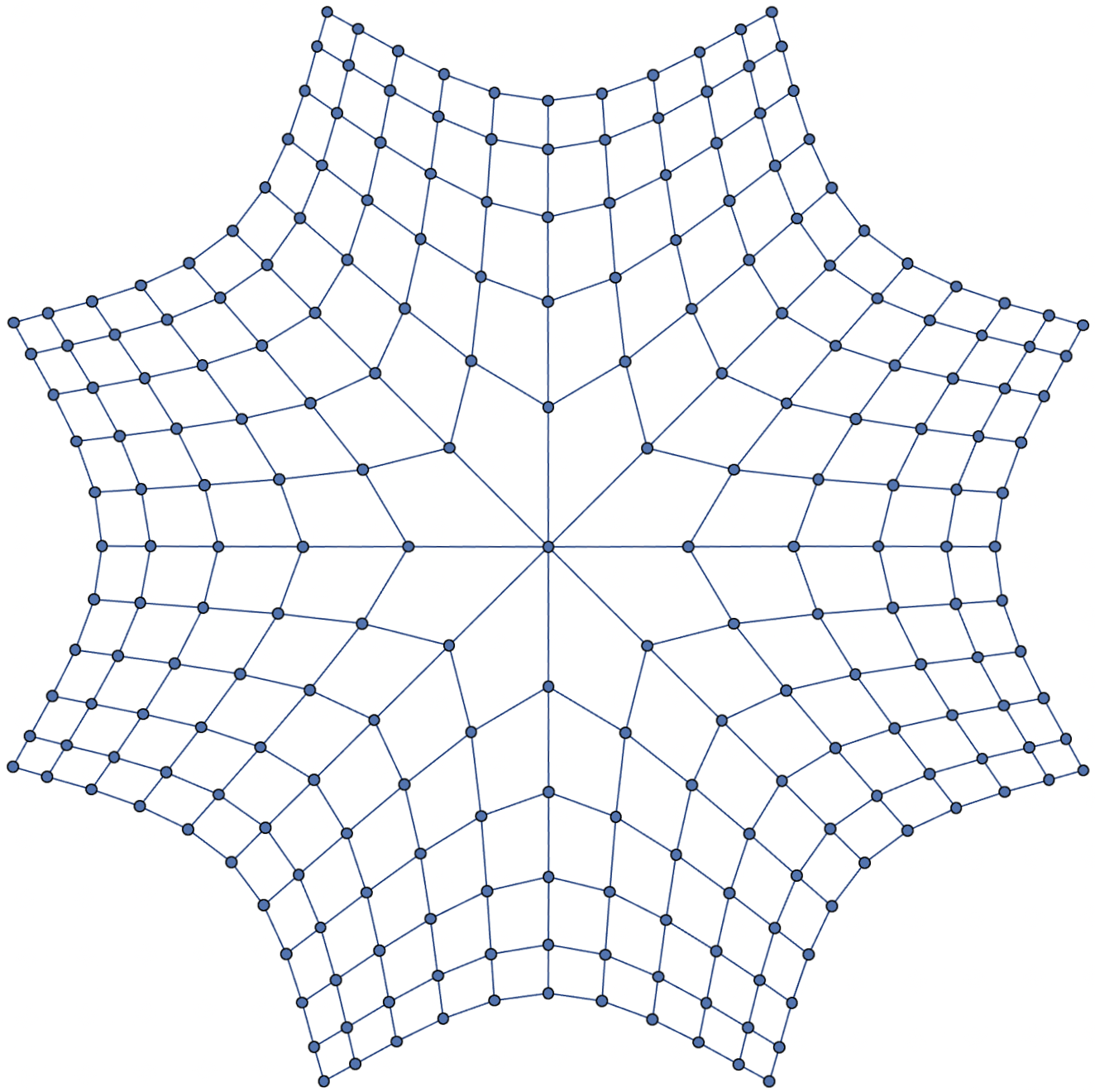}
    \caption{A $\Omega_{\text{disc}}=-2\pi$ disclination.}
    \label{fig:minus2pi_disc}
\end{figure}

In the case where $\gamma_{\OO}$ only encloses a disclination, then $\Gamma = \Omega_{\text{disc}}$, which is the disclination angle lifted to the real numbers. This definition of $\Omega_{\text{disc}}$ diverges slightly from more standard previous formulations, where $\Omega$ is defined as the angle by which a local frame (vielbein) is rotated upon being parallel transported around the defect; under such a definition, $\Omega$ is only defined modulo $2\pi$, which is problematic: A $\Omega_{\text{disc}}=-2\pi$ disclination shown in Fig.~\ref{fig:minus2pi_disc} contributes a non-trivial fractional charge $Q_W=-\SO=C/2\mod 1$ \cite{zhang2022fractional}. This issue is fixed upon treating $\Omega_{\text{disc}}$ as a lift of the disclination angle to the real numbers.

When $\gamma_{\OO}$ only encloses a boundary, the corner angle $\Omega_{\text{cor}}$ can be determined from $\Gamma$ by: 
\begin{align}
    \Omega_{\text{cor}}=\Gamma-2\pi
\end{align}
For instance, a lattice with corner angle $\Omega_{\text{cor}}=\pi/2$ is shown in Fig.~\ref{fig:lattices}(d). 

Note that the boundary in Fig.~\ref{fig:lattices}(c) also includes two corners with opposite corner angle and the total corner angle $\Omega_{\text{cor}}=0$.

\subsection{Equivalence classes of $\vec{L}_{\OO}$}
\label{sec:equivalence}

When the total corner angle for a boundary is not zero modulo $2\pi$, $\vec{L}_{\OO}$ depends on the origin $\OO$. 
If we shift $\OO$ by an integer vector $\vec{v}\in \vec{\Lambda}$, $\OO \rightarrow \OO + \vec{v}$ such that $\OO + \vec{v}$ is still a point on $\gamma_{\OO}$, then
\begin{equation}
    \vec{L}_{\OO+\vec{v}}=\vec{L}_{\OO}+(1-U(\Gamma))\vec{v},
\end{equation}
where $U(\Gamma)$ represents a counterclockwise rotation by $\Gamma$.

The shift $\OO\rightarrow\OO+\vec{v}$ does not change the high symmetry point of $\OO$; they both lie in the same maximal Wyckoff position. We define an equivalence class on $\vec{L}_{\OO}$:
\begin{equation}
    \vec{L}_{\OO}\simeq\vec{L}_{\OO}+(1-U(\Gamma))\vec{\Lambda},
\end{equation}
where $\vec{\Lambda}$ is an integer vector. Then $\vec{L}_{\OO + \vec{v}} \simeq \vec{L}_{\OO}$. Notably, $(1-U(\Gamma))\vec{\Lambda}\cdot \PO=0\mod 1$, which implies that the charge response in Eq.~\eqref{eq:charge} only depends on the equivalence class of $\vec{L}_{\OO}$ rather than its exact value \cite{manjunath2021cgt,zhang2022pol}.

We can see this equivalence in the example shown in Fig.~\ref{fig:cones}(b).
For two origins $\OO_1$ and $\OO_2$, both having MWP $\beta$ and related by an integer vector, both $\vec{L}_{\OO_1}$ and $\vec{L}_{\OO_2}$ lie in the same $[(0,0)]$ equivalence class. If $\OO_3=\alpha$, as seen in Fig.~\ref{fig:cones}(c), then $\vec{L}_{\OO_3}$ lies in the $[(0,1)]$ equivalence class instead.

We remark that a lattice with corner angle $\Omega_{\text{cor}}$ can be isometrically embedded on a cone with deficit angle $\Omega=\Omega_{\text{cor}}+2\pi$ as seen in Fig.~\ref{fig:cones}(a) This demonstrates that a lattice with a corner will reduce to a pure disclination in the limit where $\vec{L}_{\OO}$ vanishes. This equivalence between corners and disclinations will be a recurring theme and discussed in Sec.~\ref{sec:general}.

\bgroup
\def\arraystretch{1.4}
\begin{table}
    \centering
    \begin{tabular}{c|c|c}
    $\OO$&$\vec{n}_{\OO}$& $m_{\OO}$ \\
    \hline
    $\alpha$&(0,0)&1\\
    $\beta$&(1/2,1/2)&0\\
    $\gamma_1$&(0,1/2)&0\\
    $\gamma_2$&(1/2,0)&0\\
    \end{tabular}
\caption{$\vec{n}_{\OO}$ and $m_{\OO}$ in the $C_4$ symmetric unit cell}\label{table:nuc}
\end{table}
\egroup

\subsection{Definitions of $n_{W,\OO}$ and $\delta\Phi_{W,\OO}$}
\label{nwodef}

To calculate the charge associated to lattice disclinations, dislocations, boundaries, and corners, we need to account for the background charge density. Thus we need a measure of the number of unit cells in the region $W$. However, when we have lattice defects and boundaries, it is possible that the defect cores and lattice boundaries have irregular, fractional unit cells. This makes it more complicated to properly define the number of unit cells in $W$. The resolution to this is that we define a quantity $n_{W,\OO}$:
\begin{align}
\label{eq:nwo}
    n_{W,\OO} = k + \vec{L}_{\OO} \cdot \vec{n}_{\OO} + \frac{2\pi - \Gamma}{2\pi} m_{\OO} .
\end{align}
Here, $k$ is an integer, which is the number of full unit cells inside $W$. $\vec{n}_{\OO}$ is a fractional vector and $m_{\OO}$ is a fractional scalar, both of which depend only on the maximal Wyckoff position of $\OO$. Their values are tabulated in Table~\ref{table:nuc}, and are determined by fitting Eq.~\eqref{eq:charge}, \eqref{eq:nwo} to the case where the Chern number $C = 0$ and the insulating state can be fully described in terms of maximally localized Wannier functions (see App. \ref{sec:zeroC}), in which case there is an independent definition of the electric polarization. This approach is similar to the method presented in \cite{zhang2022pol}.

$n_{W,\OO}$ plays the role of an effective number of unit cells in $W$. When there are no defects or lattice boundaries in $W$, then $n_{W,\OO} = k$ is simply the integer number of unit cells in $W$, and is independent of $\OO$. However in the presence of defects and/or boundaries, the contribution of the background charge term $\nu n_{W,\OO}$ in Eq. \ref{eq:charge} necessarily depends on maximal Wyckoff position (MWP) of $\OO$. This is because the other terms in the decomposition of Eq. \ref{eq:charge}, $\mathscr{S}_{\OO} \Gamma/2\pi$, $\vec{L}_{\OO} \cdot \vec{\mathscr{P}}_{\OO}$, also depend on the MWP of $\OO$.

\begin{figure}[t]
    \centering
    \includegraphics[width=8cm]{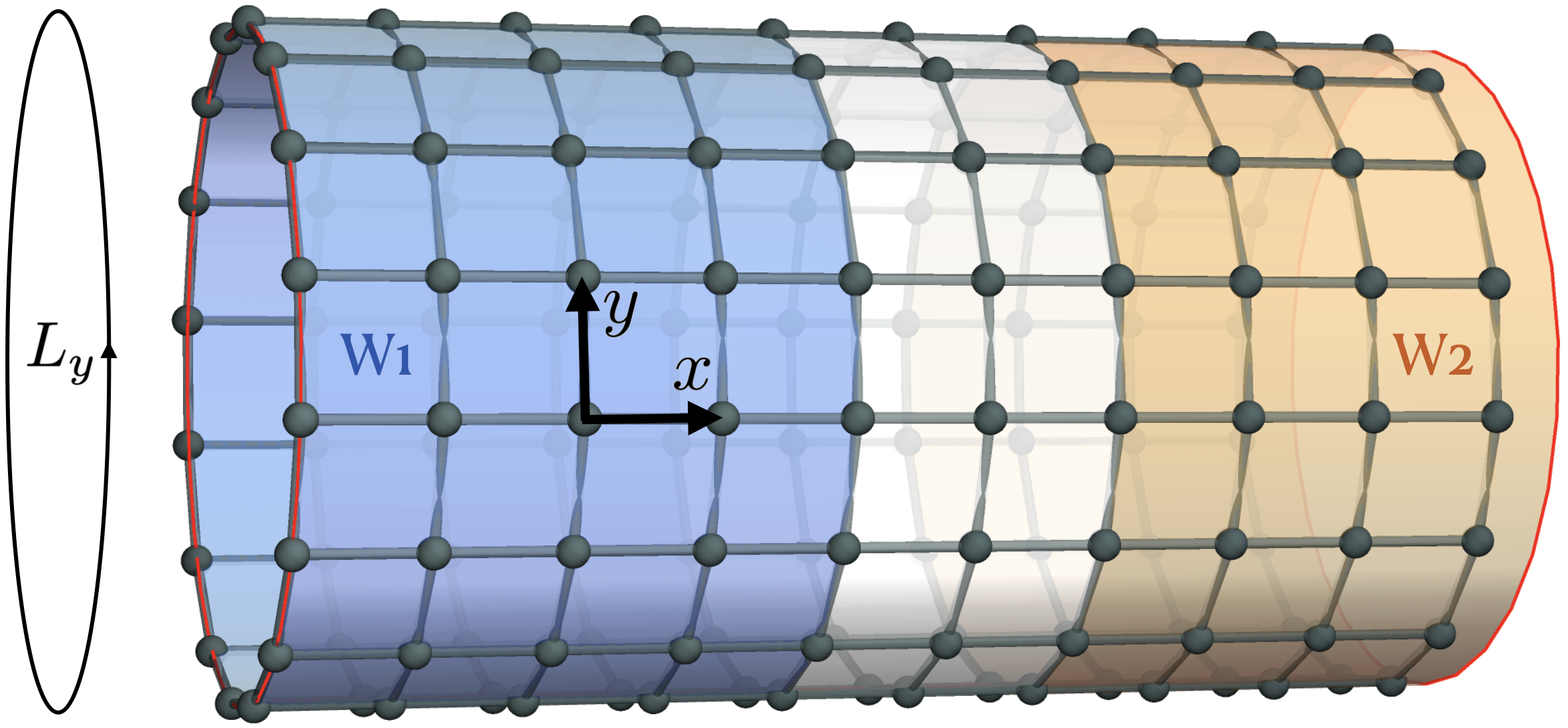}
    \caption{A cylinder with periodic boundary condition in $y$-direction. The charge calculation using the blue(orange) shaded region $W_1(W_2)$ will extract $\mathscr{P}_{\alpha,y}$($\mathscr{P}_{\beta,y}$). In this example, $n_{W_1,\beta}=4L_y$, $n_{W_2,\alpha}=(3+1/2)L_y$ is the number of unit cells inside $W_1$ and $W_2$ respectively.  The red circle is the cutoff and is not regarded as the boundary of $W$. 
    }
    \label{fig:cylinder2}
\end{figure}

\begin{figure*}[t]
    \centering
    \includegraphics[width=17.cm]{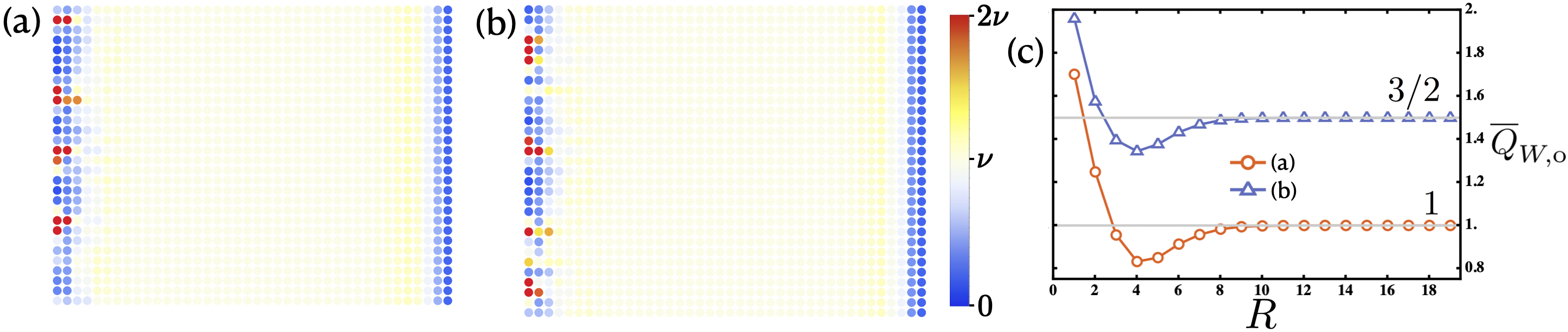}
    \caption{\textbf{(a)~(b)} The charge density at each site on a cylinder with periodic boundary condition in the y-direction. The region $W$ covers the left boundary. We have added a random on-site potential on the left boundary to show that $\overline{Q}_{W,\OO}$ is robust against perturbations. The parameters are: $\OO=\alpha$,  $C=-2$, $\phi/2\pi=\pi-\epsilon$, where $\epsilon$ is a small amount required to open the gap. $L_x\times L_y = 40\times 30$ for (a), and $L_x\times L_y = 40\times 31$ for (b). \textbf{(c)} $\overline{Q}_{W,\OO}$ converges to fractional values as $R$, the width of $W$, is large enough.}
    \label{fig:cylinder_profile}
\end{figure*}

Similarly, we define an effective excess flux in $W$ as 
\begin{align}
    \delta \Phi_{W,\OO} = \Phi_W - \phi n_{W,\OO} . 
\end{align}
$\Phi_W$ is the total flux within $W$ and $\phi$ is the flux per unit cell. $\phi n_{W,\OO}$ indicates how much flux there should be if no excess flux is inserted.

Note that if we shift $n_{W,\OO} \rightarrow n_{W,\OO} + 1$, then we can see from Eq. \ref{eq:charge} that $Q_W \rightarrow Q_W + \nu - C \phi/2\pi \mod 1 = Q_W \mod 1$, because $\kappa \equiv \nu - C \phi/2\pi$ must be an integer for any Chern insulator. Therefore, $Q_W \mod 1$ is only sensitive to $n_{W,\OO} \mod 1$. We can think of the latter as the effective fractional value of the number of unit cells in $W$.

\begin{figure*}[t]
    \centering
    \includegraphics[width=13.cm]{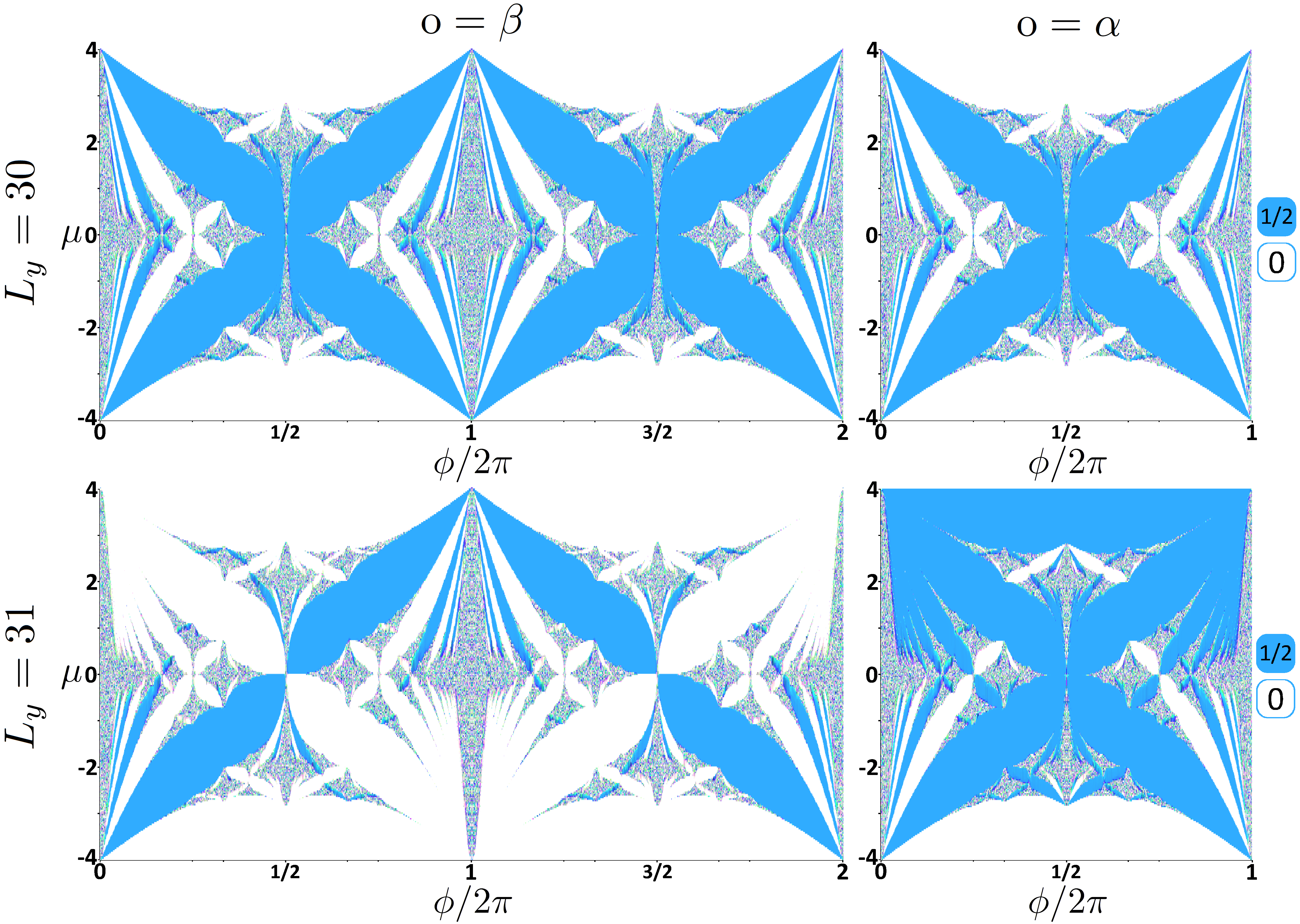}
    \caption{The full Hofstadter butterfly of the regularized charge $\overline{Q}_{W,\OO}$ for $\OO=\{\alpha,\beta\}$, $L_y=\{30,31\}$. There are 40 sites in $x$-direction. $n_{W,\OO}=15L_y(15.5L_y)$ if $\OO=\alpha(\beta)$. In lobes of higher Chern number, the correlation length is of the order of the size of the system, so finite size effects result in noise in the colored Hofstadter butterfly.}
    \label{fig:edge_charge_num}
\end{figure*}

To understand more clearly how the origin dependence of $n_{W,\OO}$ is manifested, consider the case where the lattice forms a cylinder, and $W$ encloses one of the boundaries. We can think of the lattice boundary as either having support on the vertices or the plaquette centers. Specifically, we take the lattice boundary to have support along $x = \OO_x + 1/2$, where $\OO_x$ is the $x$-component of $\OO$. With this lattice cutoff, $n_{W,\OO}$ calculated in Eq.~\eqref{eq:nwo} is the same as the area of $W$. As shown in Fig.~\ref{fig:cylinder2}, the cutoff, depicted as the red cycle, is vertically crossing the sites when $\OO=\alpha$, and the cutoff is vertically crossing the plaquette center when $\OO=\beta$. The fractional number of unit cells per unit length can then be visualized in terms of the area between the red cycle and the closest unit cell boundary. 

It is important to clarify that the choice of the boundary cutoff (red cycle in Fig. \ref{fig:cylinder2}) should not be conflated with the smooth and rough boundaries of the lattice. The latter refers to the lattice configuration at the boundary, while the former is a theoretical tool to define the total number of unit cells $n_{W,\OO}$, without imposing any constraint on the lattice configuration at the boundary. Our numerics confirms that the lattice configuration at the boundary does not affect the calculation of the $\SO$ and $\PO$ invariants, regardless of whether the boundary is smooth, rough or even more exotic, such as breaking translation symmetry in the $y$-direction.

\section{Charge calculation}\label{sec:charge}

In this section we numerically verify Eq. \ref{eq:charge}, focusing on the polarization and discrete shift contributions separately by first studying boundary charge on a cylinder and second by studying corner contributions to the charge. 

\subsection{Edge charge and quantized electric polarization}\label{sec:edgeC}

In this section, we focus on the explicit calculation of the electric polarization $\PO$ using boundary charge. We consider the Hofstadter model defined on a cylindrical geometry with periodic boundary conditions along the $y$-axis and open boundary conditions along the $x$-axis, as shown in Fig. \ref{fig:cylinder2}.

The Hofstadter Hamiltonian is
\begin{align}
    H_{\text{cylinder}} = -\sum_{\langle ij\rangle} t_{ij} c_i^\dagger c_j + \text{h.c.}, 
\end{align}
where the nearest neighbour hopping $t_{ij}\equiv t e^{i A_{ij}}$ defines the vector potential $A_{ij}$ with $\phi$ flux per plaquette.

We define a large region $W$ which fully encloses one of the boundaries of the cylinder shown in Fig.~\ref{fig:cylinder2}. In this example, the vectorized edge length is $\vec{L}_{\OO} = (0,L_y)$, $\Gamma = 2\pi$ and the corner angle $\Omega_{\text{cor}}=0$. 

We empirically find that $Q_{W}$ obeys
\begin{equation}\label{eq:cylinderC}
    Q_{W} = \frac{C}{2} +\vec{L}_{\OO}\cdot\vec{\mathscr{P}}_{\OO} + \nu n_{W,\OO}+\frac{C\delta\Phi_{W,\OO}}{2\pi}\mod 1.
\end{equation}
This agrees with Eq.~\eqref{eq:charge} after using the relation $\mathscr{S}_{\OO} \mod 1 = C/2 \mod 1$ \cite{zhang2022fractional}. 
 To show this, we set $\delta \Phi_{W,\OO} = 0$ in the Hamiltonian and present the explicit numerical data for the regularized charge after subtracting the background contribution:
\begin{align}
\label{eq:QWbar}
    \overline{Q}_{W,\OO}\equiv Q_{W}-\nu n_{W,\OO}.
\end{align}
In Fig.~\ref{fig:cylinder_profile}, we show the charge profile on a cylinder where $\overline{Q}_{W,\OO}$ converges to the predicted fraction for a large enough $W$. The quantization is shown to persist even in the presence of random on-site perturbations along the boundary as shown in Fig.~\ref{fig:cylinder_profile}. In Fig. \ref{fig:edge_charge_num}, we show the full colored Hofstadter butterfly for $\overline{Q}_{W,\OO}$, where $\mathscr{P}_{\OO,y}$ can be extracted. We find that the extracted $\mathscr{P}_{\OO,y}$ agrees with $\mathscr{P}_{\OO,y}$ calculated using other methods in \cite{zhang2022pol}. 

\begin{figure*}[t]
    \centering
    \includegraphics[width=17.5cm]{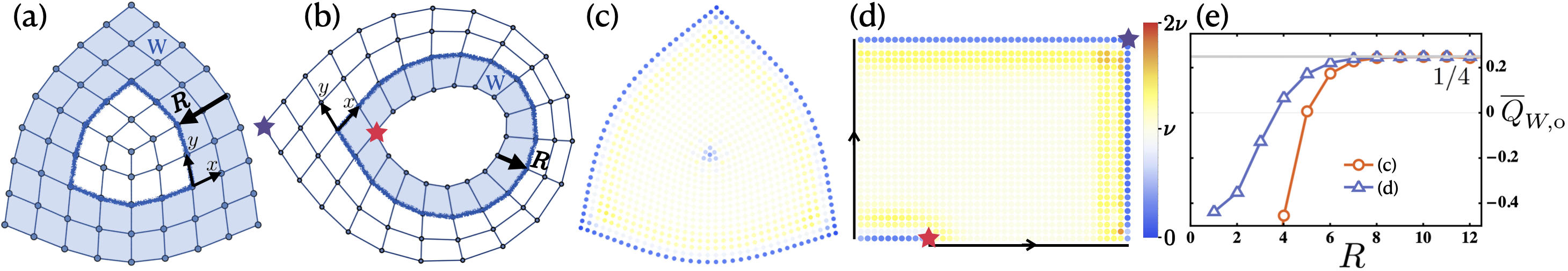}
    \caption{\textbf{(a)} Region $W$ covering the outer boundary of a $\pi/2$ disclination. In this example $\Gamma=7\pi/2$, $\vec{L}_{\beta}\simeq[(0,0)]$, $\vec{L}_{\alpha}\simeq[(1,0)]$,
    $n_{W,\beta}=36$,
    $n_{W,\beta}=36-3/4$.
    \textbf{(b)} A ribbon geometry where $W$ encloses a corner with $\Omega_{\text{cor}}=-\frac{\pi}{2}$ and a full edge. This lattice is effectively a top-down projection of the lattice in Fig.~\ref{fig:cones}. Although visually distorted, each unit cell is regarded as a perfect square. 
    In this example $\Gamma=3\pi/2$. $\vec{L}_{\beta}\simeq[(0,0)]$, $\vec{L}_{\beta}\simeq[(1,0)]$, $n_{W,\beta}=17$, $n_{W,\alpha}=17+1/4$. \textbf{(c)-(d)} charge profile of the geometry in (a) and (b). In (d) we glue the two boundary along the arrow direction to obtain the ribbon geometry. The inner corner and outer corners are labeled with stars. The parameters are $C=2$, $\phi=\pi+\epsilon$ $\OO=\beta$. \textbf{(e)} $\overline{Q}_{W,\OO}$ converges to fractional values for a large enough $R$. }
    \label{fig:corner_charge_num}
\end{figure*}

A naive interpretation of a two-dimensional polarization would be that it specifies a charge per unit length along the boundary. However such a definition is complicated because one needs to disentangle the boundary and the bulk charge, and furthermore it is not robust to perturbations along the boundary that break translational symmetry. Our results demonstrate that one can generically define a boundary charge and obtain a precise definition of $\vec{\mathscr{P}}_{\OO}$ that is robust to random perturbations along the boundary in terms of the oscillatory system size dependence of $Q_{W} \mod 1$.

As found in \cite{zhang2022pol}, an oscillatory $L_y$-dependent response also appears when we view the cylinder as an effective 1d system and compute the effective 1d polarization along the $x$ direction. Indeed $\overline{Q}_{W,\OO}$, which is the boundary contribution of the charge, effectively specifies a polarization of the 1d system through a net dipole moment along the $x$-direction. The results here provide an alternative way to extract $\vec{\mathscr{P}}_{\OO}$ through a careful definition of the boundary charge $\overline{Q}_{W,\OO}$. 

\subsection{Corner charge}\label{sec:cornerC}

\begin{figure*}[t]
    \centering
    \includegraphics[width=18cm]{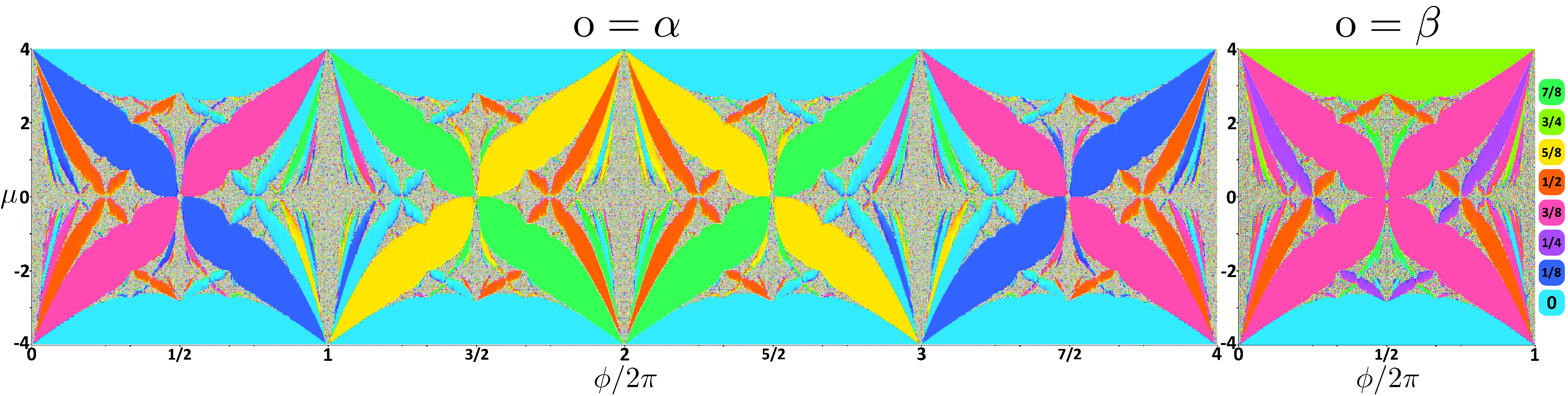}
    \caption{The full Hofstadter butterfly of the regularized ribbon charge
$\overline{Q}_{W,\OO}$ for $\OO=\{\alpha,\beta\}$. The total number of sites at the inner boundary is 10, 11 for $\OO=\beta,\alpha$ and the width of the ribbon is 30. In both cases $\vec{L}_{\OO}\in [(0,0)]$}
    \label{fig:ribbonC}
\end{figure*}

We now outline the methodology for calculating the corner charge in the square lattice Hofstadter model. 
A corner is inherently conjoined with 2 edges. Furthermore, the presence of gapless chiral edge states implies that we cannot directly consider the charge near a corner. Rather, we must consider a region containing a full boundary. 
We can then isolate the contribution from the corners by taking into account the contribution from the polarization and other bulk contributions. 

A familiar example of a corner geometry is shown in Fig.~\ref{fig:corner_charge_num}(a) which is the outer boundary of a lattice with a disclination, where $\Gamma=\frac{7\pi}{2}$. 
Another example is shown in Fig.~\ref{fig:corner_charge_num}(b) where the region $W$ contains only one corner with the minimal corner angle $\Omega_{\text{cor}}=-\frac{\pi}{2}$, $\Gamma=\frac{3\pi}{2}$.

We can numerically calculate $Q_W$ for Fig.\ref{fig:corner_charge_num}(a)(b). They follow:
\begin{align}\label{eq:cornerC0}
    Q_{W} = \frac{C}{2} + \frac{3}{4}\mathscr{S}_{\OO} + \vec{L}_{\OO}\cdot \vec{\mathscr{P}}_{\OO} + \nu n_{W,\OO}+\frac{C\delta\Phi_{W,\OO}}{2\pi}\mod 1,
\end{align}
\begin{align}\label{eq:cornerC}
    Q_{W} = \frac{C}{2} - \frac{1}{4}\mathscr{S}_{\OO} + \vec{L}_{\OO}\cdot \vec{\mathscr{P}}_{\OO} + \nu n_{W,\OO}+\frac{C\delta\Phi_{W,\OO}}{2\pi}\mod 1,
\end{align}
which agrees with Eq.~\eqref{eq:charge} after using the relation $\mathscr{S}_{\OO} \mod 1 = C \mod 1$. Similar to the edge charge calculation in the preceding section, we consider systems where $\delta\Phi_{W,\OO}=0$, and define $\overline{Q}_{W,\OO}\equiv Q_{W}-\nu_{W,\OO}$. We show in Fig.~\ref{fig:corner_charge_num}(c,d,e) the charge profile of $\overline{Q}_{W,\OO}$ which converges to the predicted fraction for a large enough $W$.

In Fig. \ref{fig:corner_charge_num}(a), one can consider the complement of $W$, $\overline{W}$, which has the same boundary $\partial \overline{W}\equiv \partial W$. $\overline{W}$ characterizes the familiar disclination charge. Since the total charge over the manifold is a integer, we can obtain the boundary charge by $Q_W=-Q_{\overline{W}}\mod 1$. Such procedure gives the same result as in Eq.~\eqref{eq:cornerC0}.

We additionally show the full Hofstadter butterfly of $\overline{Q}_{W,\OO}$ for the Fig.~\ref{fig:corner_charge_num}(b) geometry in Fig.~\ref{fig:ribbonC}, where we have considered geometries where $\vec{L}_{\OO}\simeq [(0,0)]$ such that the $\PO$ contribution vanishes.

It is worth noting that since $\SO\mod 1=C/2\mod 1$, Eq.~\eqref{eq:cornerC} can be reformulated as 
\begin{equation}\label{eq:cornerC2}
    Q_{W} = \vec{L}_{\OO}\cdot \vec{\mathscr{P}}_{\OO}+\frac{3}{4}\mathscr{S}_{\OO} +\nu n_{W,\OO}+\frac{C\delta\Phi_{W,\OO}}{2\pi}\mod 1.
\end{equation}
This alternative expression offers a different interpretation: $Q_W$ represents the charge of a impure disclination with burgers vector $\vec{b}_{\OO}=\vec{L}_{\OO}$ and disclination angle $\Omega_{\text{disc}}=2\pi+\Omega_{\text{cor}}=3\pi/2$. In the $\vec{L}_{\OO} = 0$ limit the ribbon geometry reduces to a pure disclination with origin $\OO$ and $\Omega_{\text{disc}}=3\pi/2$ as demonstrated in Fig.~\ref{fig:cones}(a). This suggests the existence of a generalized framework for understanding crystalline defects and boundaries, which we explore in Sec. \ref{sec:general}.

\section{Equivalence between boundaries and bulk defects}\label{sec:general}

We now present a more in-depth understanding of the equivalence between edges and dislocations, and corners and disclinations. In Sec.~\ref{sec:cornerC} we showed an example where a $\Omega_{\text{cor}}=-\frac{\pi}{2}$ corner with vectorized edge length $\vec{L}_{\OO}$ is equivalent to a $\Omega=\frac{3\pi}{2}$ pure disclination in the limit $\vec{L}_{\OO}=(0,0)$.
This statement can be further generalized as follows. 

Consider an orientable 2-manifold $\mathcal{M}$ with genus $g$ and number of boundaries $n_{\text{boundary}}$. The Gauss-Bonnet theorem is given by:
\begin{equation}\label{eq:gb}
    \chi=\frac{1}{2\pi}\int_{\mathcal{M}}RdA + \frac{1}{2\pi} \int_{\partial \mathcal{M}} Kds
\end{equation}
where $\chi=2-2g-n_{\text{boundary}}$ is the Euler characteristic of the manifold $\mathcal{M}$. $R$ is the Gaussian curvature, and $dA$ is an area element; $K$ is the geodesic curvature on the boundary $\partial \mathcal{M}$, and $ds$ is a line element.  The geodesic curvature $K$ is formally defined as the norm of the covariant derivative $K=||D\hat{T}/ds||$ , where $\hat{T}$ is the unit tangent vector along the boundary. 

Now consider a lattice which in the bulk contains lattice disclinations with disclination angle $\Omega_{\text{disc},i}$ and on the boundary contains corners with corner angles $\Omega_{\text{cor},j}$. Here $i$ and $j$ label the disclinations and corners respectively. The total disclination and corner angles are $\Omega_{\text{disc}} = \sum_i \Omega_{\text{disc},i}$ and $\Omega_{\text{cor}} = \sum_j \Omega_{\text{cor},j}$. 
A quantum many-body system defined on such a lattice will be described at low energies by a quantum field theory defined on a spatial manifold $\mathcal{M}$, where the lattice disclinations and corners can be modeled in the continuum geometry by delta function sources of bulk and boundary curvature. That is,
\begin{align}
\int_{\mathcal{M}} R d A = \sum_{i} \Omega_{\text{disc},i}, \;\;\; \int_{\partial \mathcal{M}} K ds = \sum_j \Omega_{\text{cor},j}.
\end{align}
Eq.~\eqref{eq:gb} can then be reformulated as
\begin{equation}\label{eq:euler}
    \frac{\sum_{i}\Omega_{\text{disc},i}}{2\pi}+\frac{\sum_{j}\Omega_{\text{cor},j}}{2\pi}=2-2g-n_{\text{boundary}}.
\end{equation}

A corollary of Eq.~\eqref{eq:euler} is that we can choose one of two different ways to label a boundary.
\begin{enumerate}
    \item A boundary with $n_{\text{bounday}}=1$ and total edge curvature $\Omega_{\text{cor}}$, vectorized edge length $\vec{L}_{\OO}$,
    \item A (impure) disclination with disclination angle $\Omega_{\text{disc}}=\Omega_{\text{cor}}+2\pi$ and Burgers vector $\vec{b}_{\OO}=\vec{L}_{\OO}$, and there is no boundary at all, $n_{\text{boundary}}=0$ .
\end{enumerate}
This is consistent with the intuition from Sec. \ref{sec:main_result} in that two sets of lattice defects/boundaries contribute equivalently to the charge mod $1$ if their $\Gamma$ and $\vec{L}_{\OO}$ are the same.

By this corollary, in the charge calculation we could choose to separate the contributions from bulk lattice defects (dislocations and discliantions) and the lattice boundaries (edges and corners) instead of packaging all of the information in $\Gamma$ and $\vec{L}_{\OO}$. Eq.~\eqref{eq:charge} can be equivalently written as:

\begin{align}\label{eq:separate_response}
    \nonumber Q_{W} &= \frac{C}{2}n_{\text{boundary}}+\frac{\Omega_{\text{disc}}}{2\pi}\mathscr{S}_{\OO} +  \frac{\Omega_{\text{cor}}}{2\pi}\mathscr{S}_{\OO}+\vec{b}_{\OO}\cdot\vec{\mathscr{P}}_{\OO}\\
    &+\vec{L}_{\OO}\cdot\vec{\mathscr{P}}_{\OO}+\nu n_{W,\OO}+C\frac{\delta\Phi_{W,\OO}}{2\pi}\mod 1
\end{align}
where we have used the relation $\SO\mod 1=C/2\mod 1$. Here $\vec{b}_{\OO}$ is the total Burgers vector of the lattice disclinations and dislocations in $W$, while $\vec{L}_{\OO}$ is the vectorized edge length along any lattice boundaries in $W$.

To intuitively understand the $\frac{C}{2}$ term, consider a cylinder constructed by removing the opposite faces of a rectangular cuboid. 
Each face of the cuboid is conjoined with four $\frac{\pi}{2}$ disclinations. The action of removing one face can be considered local as it is far from the boundary of region $W$. Therefore, one end of the cylinder can be considered to have total disclination angle $2\pi$, resulting in an extra $\SO\mod 1$ contribution to $Q_{W}$ in Eq.~\eqref{eq:separate_response}. Using the proven relation $\SO\mod 1=C/2 \mod 1$ \cite{zhang2022fractional}, attributing the hole of the cylinder as a $2\pi$ disclination recovers the $\frac{C}{2}$ contribution to $Q_{W}$ on the cylinder.

\begin{figure}[t]
    \centering
    \includegraphics[width=4.5cm]{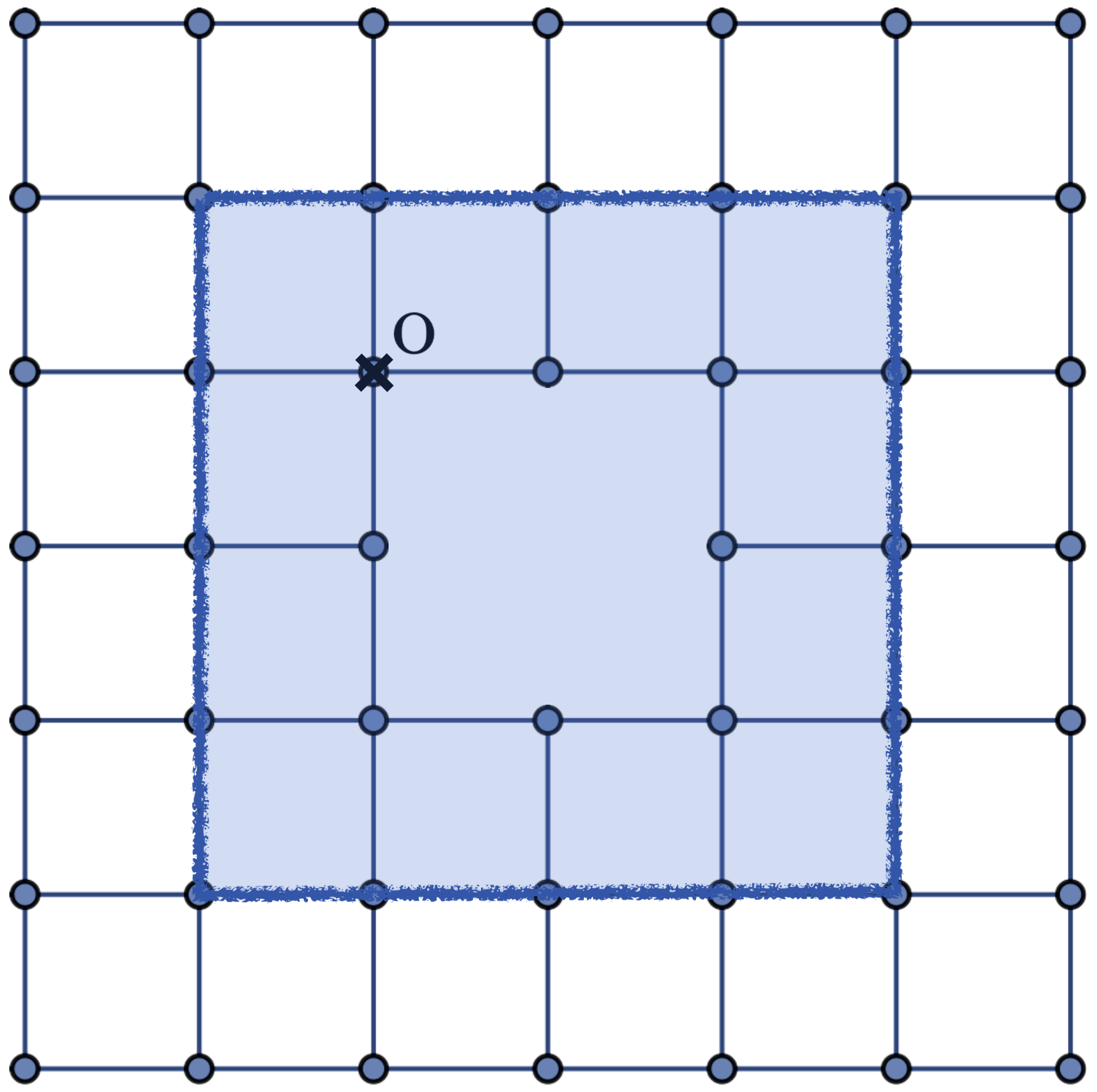}
    \caption{The region $W$ covering a trivial defect which is created by removing a site.}
    \label{fig:trivial}
\end{figure}

A different way of understanding the $\frac{C}{2}$ contribution is that this term is necessary to regularize the charge such that a trivial defect will have $Q_W=0 \mod 1$. Consider a trivial defect constructed by removing a site, setting $\OO=\beta$, as shown in Fig.~\ref{fig:trivial}. 
Before inserting this defect, $Q_W=16\nu \mod 1$ only receives contributions from the $\nu n_{W,\OO}$ term.
Since this trivial defect is a local modification of the system, $Q_W$ should not change modulo 1 after inserting the defect. We can model this defect as having one boundary, four corners with total corner angle $\Omega_{\text{cor}}=-2\pi$, an edge length $\vec{L}_{\beta}=(0,0)$, and four fewer unit cells as compared to the clean lattice while maintaining the same total flux $\Phi_W$. The charge response, according to Eq.~\ref{eq:charge}, reads 

\begin{align}\label{eq:trivial_defect}
    \nonumber Q_{W} &= \frac{C}{2} + \vec{0} \cdot\vec{\mathscr{P}}_{\beta} - \mathscr{S}_{\beta}\frac{2\pi}{2\pi}   + \nu (16-4) +\frac{4C\phi}{2\pi}\mod 1\\
    \nonumber &= \frac{C}{2} - \mathscr{S}_{\beta} + 16\nu -4\kappa \mod 1\\
    &= 16\nu \mod 1,
\end{align}

Here, we have again used the proven relation $\mathscr{S_{\OO}}\mod 1=\frac{C}{2} \mod 1$. The charge before and after inserting the trivial defect is both $16\nu \mod 1$ as expected since a trivial defect contributes integer charge. This means that for each boundary within $W$, there should be a  $\frac{C}{2}$ charge which is crucial for regularizing $Q_W$. It is easy to check that with $\OO=\alpha$, $Q_W=16\nu \mod 1$ as well and the same argument applies.

The discussion above shows that another way to view lattice boundaries with corners is entirely using the framework of disclinations and Burgers vectors. To use this perspective, we shift 

\begin{align}
    \nonumber&\{\Omega_{\text{disc}},\Omega_{\text{cor}},n_{\text{boundary}}\}\\
    \rightarrow&\{\Omega_{\text{disc}}+\Omega_{\text{cor}}+2\pi n_{\text{boundary}},0,0\},
\end{align}
and treat $\vec{L}_{\OO} = \vec{b}_{\OO}$ as a Burgers vector. This is the perspective that we will take in connecting our results to the topological effective action derived in \cite{manjunath2021cgt,manjunath2020FQH,zhang2022fractional,zhang2022pol} and reviewed below.

\section{Topological crystalline gauge theory description}\label{sec:field}

In this section, we discuss how the results of the preceding sections can be understood through the framework of topological quantum field theory. Much of this section is a review of results presented in \cite{manjunath2021cgt,manjunath2020FQH,zhang2022fractional,zhang2022pol}.

Refs. \cite{manjunath2021cgt,manjunath2020FQH} developed the topological effective action for topological phases of bosons with symmetry group $G = U(1) \times_\phi [\mathbb{Z}^2 \rtimes \mathbb{Z}_M]$. The results were then extended to invertible fermionic topological phases in \cite{zhang2022fractional,zhang2022pol}, using the general theory of invertible fermionic topological phases of \cite{barkeshli2021invertible}. 
We proceed by introducing the background $U(1)$ gauge field $A$, a background $\mathbb{Z}_M$ gauge field $\omega_\OO$ associated with $\mathbb{Z}_M$ point group rotations about $\OO$, and a background $\mathbb{Z}^2$ gauge field $\vec{R}$. Together $(A, \omega_\OO, \vec{R})$ define a background gauge field for the $G$ symmetry. $\omega_\OO$ and $\vec{R}$ are referred to as crystalline gauge fields. In particular, $\omega_\OO$ is referred to as the rotation gauge field, while $\vec{R}$ is referred to as the translation gauge field. Their holonomies encode geometrical properties of the lattice \cite{manjunath2021cgt}. $\int_W d \omega_\OO$ encodes the disclination angle in the region $W$, while $\int_W d \vec{R}$ encodes the Burgers vector in the region $W$. For convenience we drop the subscript on $\omega_\OO$ below.  

Mathematically, it is helpful to begin in a simplicial formulation. We start with a 3-dimensional space-time manifold $\mathcal{N}$ and triangulate it. Then, $A_{ij} \in \mathbb{R}$, $\omega_{ij} \in \frac{2\pi}{4} \mathbb{Z}$ and $\vec{R}_{ij}$ are defined on 1-simplices $(ij)$ of the triangulation, where $i,j$ label 0-simplices (vertices). That is, mathematically they are 1-cochains defined on the triangulation. Note that $A_{ij}$ and $\omega_{ij}$ are technically lifts of the $U(1)$ and $\mathbb{Z}_4$ gauge fields to $\mathbb{R}$ and $\frac{2\pi}{4}\mathbb{Z}$, respectively, which is important for defining the topological action. The action is then independent of the choice of lift. 

To obtain a topological action for invertible bosonic topological phases, we pick a cohomology class $[\nu_3] \in H^3(BG, \mathbb{R}/\mathbb{Z})$, where $BG$ is the classifying space \cite{dijkgraaf1990,Chen2013}. The gauge field can be interpreted as a map from the space-time manifold to the classifying space. Thus, assuming that $A, \omega, \vec{R}$ are flat gauge fields, we can use them to pull back the 3-cocycle $\nu_3$ to a 3-cocycle defined on the space-time manifold. On a closed space-time manifold, this leads \cite{manjunath2021cgt} to a topologically invariant action 
$S = \int_\mathcal{\mathcal{N}} \mathcal{L}$, with Lagrangian density  
\begin{align}
\label{action_simplicial}
    \mathcal{L} = &\frac{C}{4\pi} A \cup d A + \frac{\mathscr{S}_{\OO}}{2\pi} A \cup d \omega + \frac{\vec{\mathscr{P}}_{\OO}}{2\pi} \cdot A \cup d \vec{R} 
    \nonumber \\
    &+ \frac{\kappa}{2\pi} A \cup A_{XY} + \cdots ,
\end{align}
where $\cdots$ includes topological terms not involving $A$, which do not concern us in this paper. Here $\cup$ denotes the cup product and $d$ denotes the coboundary operation. Here $A_{XY}$ is a 2-cochain, whose explicit formula in terms of $\vec{R}$, $\omega$ is given in \cite{manjunath2021cgt}. If we take $W$ to be a spatial region, $\int_W A_{XY}$ represents the number of unit cells in $W$.\footnote{Note that the number of unit cells in $W$ is a property of the lattice model, not the triangulation of $\mathcal{N}$. The latter is a mathematical tool to define the topological action.}

In order to use more familiar notation, it is convenient to recast the above action in a continuum formulation. In this formulation, $A$, $\omega$, and $\vec{R}$ are real-valued differential forms, and the action is written in the continuum using differentials and wedge products:
\begin{align}
\label{top_action}
    \mathcal{L} = &\frac{C}{4\pi} A \wedge d A  + \frac{\mathscr{S}_{\OO}}{2\pi} A \wedge d \omega + \frac{\vec{\mathscr{P}}_{\OO}}{2\pi} \cdot A \wedge d\vec{R} 
    \nonumber \\ 
    & + \frac{\kappa}{2\pi} A \wedge A_{XY} + \cdots.
\end{align}
The continuum versions of the gauge fields are defined such that integrating over a simplex gives the corresponding quantity in the simplicial formulation. 

As explained in \cite{manjunath2021cgt}, the rotation gauge field $\omega$ is closely related to the $SO(2)$ spin connection on the spatial manifold on which the system is defined. A lattice system with a disclination is expected to be described at long wavelengths by a quantum field theory defined on a manifold where the disclination corresponds to a conical singularity, that is, a delta function source of curvature. Thus, if we split our space-time manifold into space and time, $\mathcal{N} = \mathcal{M} \times S^1$, the space $\mathcal{M}$ has conical singularities at the locations corresponding to the lattice disclinations. Therefore, the rotation gauge flux $d \omega$ is also equal to the geometric curvature. Nevertheless, it is important to distinguish the $SO(2)$ spin connection from the $\mathbb{Z}_M$ rotation gauge field $\omega$, because the discrete character of the rotation gauge field implies different possibilities for the classification of invariants and topological terms. The topological terms presented here can then be viewed as a discrete cousin of analogous terms that appear in the geometric response of continuum quantum Hall systems \cite{Wen1992shift,Gromov2014,Gromov2015}.

In the case of invertible fermionic phases, such as Chern insulators, which are the focus of this paper, we first use the classification of \cite{barkeshli2021invertible}. For the case of invertible fermionic phases with $U(1)^f$ symmetry,\footnote{$U(1)^f$ denotes the group $U(1)$, where the order-2 element corresponds to fermion parity} invertible phases can be classified by $(c_-, n_2, \nu_3)$, where $c_- \in \mathbb{Z}$ is the chiral central charge, $n_2$ is a $\mathbb{Z}_2$-valued $2$-cochain on $BG$, while $\nu_3$ is an $\mathbb{R}/\mathbb{Z}$-valued $3$-cochain on $BG$. As in the bosonic case, the topological action then corresponds to the pullback of $\nu_3$ onto the space-time manifold $\mathcal{N}$, and results in the same Lagrangian density as in the bosonic case, Eq. \ref{action_simplicial},\ref{top_action}. The only difference is the quantization of the invariants $C, \mathscr{S}_{\OO}$. In the bosonic case, $C$ must be an even integer and $\mathscr{S}_\OO$ is an integer. In the fermionic case $C$ is any integer, $\mathscr{S}_{\OO}$ can be half-integer, and we have the identity $\mathscr{S}_{\OO} = C/2 \mod 1$\cite{zhang2022fractional}. 

We can use the topological action to obtain the charge in a region $W$:
\begin{align}
    Q_W  =  \int_W \frac{\delta \mathcal{L}}{\delta A_0} 
    = C \frac{\Phi_W}{2\pi} + \mathscr{S}_{\OO} \frac{\Omega_{\text{disc}}}{2\pi} + \vec{\mathscr{P}}_{\OO} \cdot \vec{b} + \kappa n_{W} 
\end{align}
Here $\Phi_W = \int_W d A$, $\Omega_{\text{disc}} = \int_W d \omega$, $\vec{b} = \frac{1}{2\pi}\int_W d \vec{R}$, and $n_W = \frac{1}{2\pi}\int_W A_{XY}$, and $A_0$ is the time-component of $A$. 

So far, the topological action is defined for flat gauge fields. This translates  to the requirement that for any region $W$, $\Phi_W/2\pi$, $\Omega_{\text{disc}}/2\pi$, $(1 - U(2\pi/M))^{-1} \vec{b}$, and $n_W$ are integer-valued \cite{manjunath2021cgt}. While the topological action was derived for flat gauge fields, we will use it to deduce the response of the system to non-flat configurations of the gauge fields, which physically means that the region $W$ contains non-trivial lattice disclinations and dislocations. This leads to complications where $n_W$ is no longer well-defined in the topological gauge theory and the Burgers vector $\vec{b}$ necessarily depends on a choice of high symmetry point $\OO$ when disclinations are present in the system. 

Therefore, motivated by the topological field theory result, we consider the prediction:
\begin{align}
    Q_W & = C \frac{\Phi_W}{2\pi} + \mathscr{S}_{\OO} \frac{\Omega_{\text{disc}}}{2\pi} + \vec{\mathscr{P}}_{\OO} \cdot \vec{b}_{\OO} + \kappa n_{W,\OO}  \mod 1
    \nonumber \\
    & = C \frac{\delta \Phi_{W,\OO}}{2\pi} + \mathscr{S}_{\OO} \frac{\Omega_{\text{disc}}}{2\pi} + \vec{\mathscr{P}}_{\OO} \cdot \vec{b}_{\OO} + \nu n_{W,\OO} \mod 1,
\end{align}
where in the second line we have used that the charge per unit cell satisfies $\nu = \kappa + C \phi/2\pi$, $\phi$ is the flux per unit cell, and we defined $\delta \Phi_{W,\OO} \equiv \Phi_W - \phi n_{W,\OO}$. This equation incorporates the fact that the Burgers vector $\vec{b}_{\OO}$ must generically depend on a choice of high symmetry point $\OO$. The modular reduction incorporates the charge quantization, $\Phi_W \sim \Phi_W + 2\pi$ by a large gauge transformation, and also the fact that non-topological effects like local potentials can change the charge in a given region by an integer. Empirically we find that this equation does successfully account for the charge in the region $W$, provided that $n_{W,\OO}$ is suitably defined, as discussed in Sec. \ref{nwodef}. 

The discussion above explicitly accounted for boundaries and corners in $W$ by treating them using the formalism of disclinations and dislocations. An alternative way to proceed would be to explicitly derive a boundary effective action, and then use it to compute the charge response at the boundary. Such a method was pursued for continuum quantum Hall systems and the Wen-Zee term in \cite{gromov2016boundary} and used to understand the filling anomaly in higher order topological insulators in \cite{rao2023effective}. We leave it to future work to explicitly derive a general boundary effective action for the topological crystalline gauge theory used in this paper.

\section{Application to quadrupole and higher-order topological insulators}\label{sec:HOTI}

\begin{figure}[t]
    \centering
    \includegraphics[width=3cm]{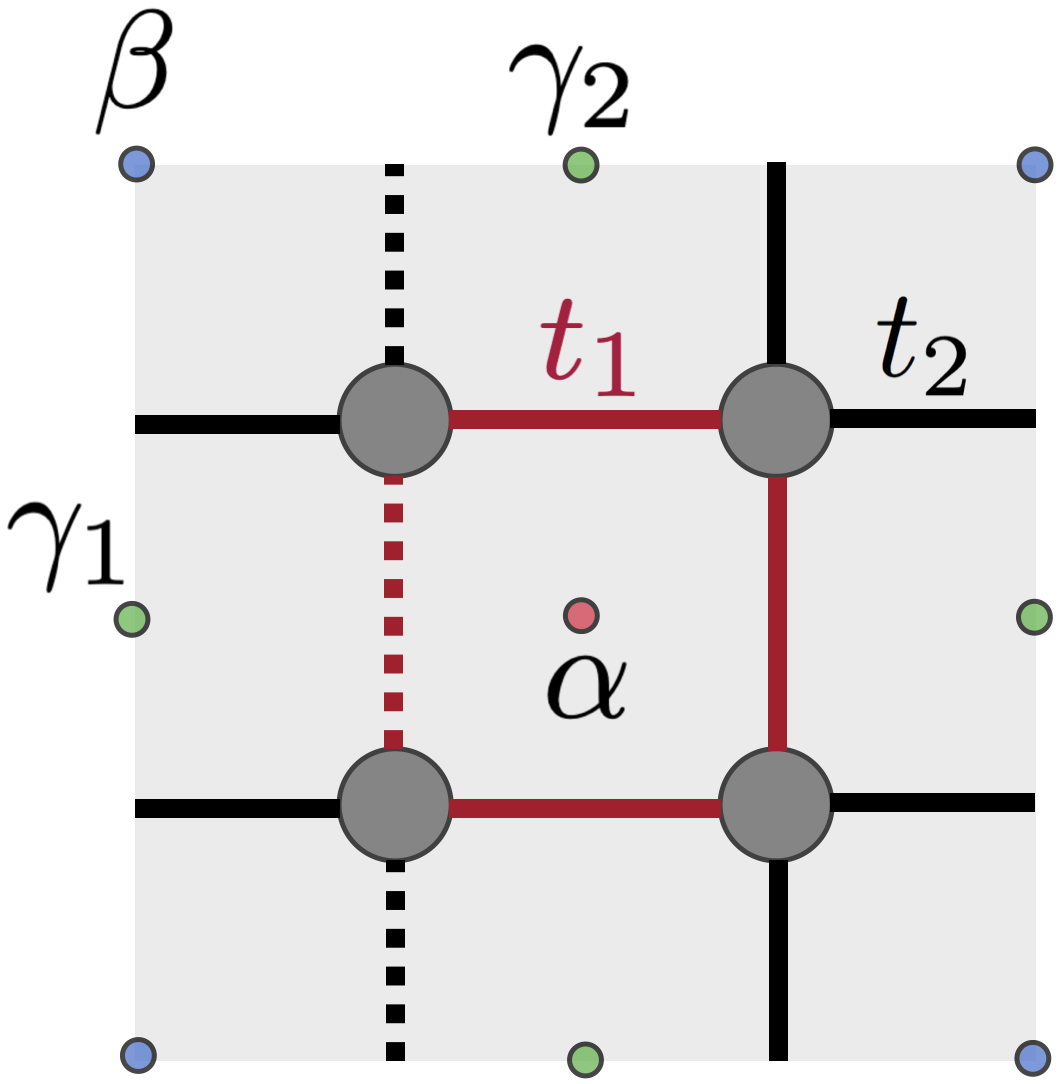}
    \caption{Unit cell of the QI model, grey dots represent sites, red links represent hoppings with amplitude $t_1$ black links connecting adjacent unit cells represent hoppings with amplitude $t_2$. Dotted lines have negative hopping amplitude. This inserts $\pi$ flux per plaquette. The colored dots are MWPs $\alpha,\beta,\gamma$. The sites are not at the MWPs in this example}
    \label{fig:qiUnit}
\end{figure}

\begin{figure*}[t]
    \centering
    \includegraphics[width=17.5cm]{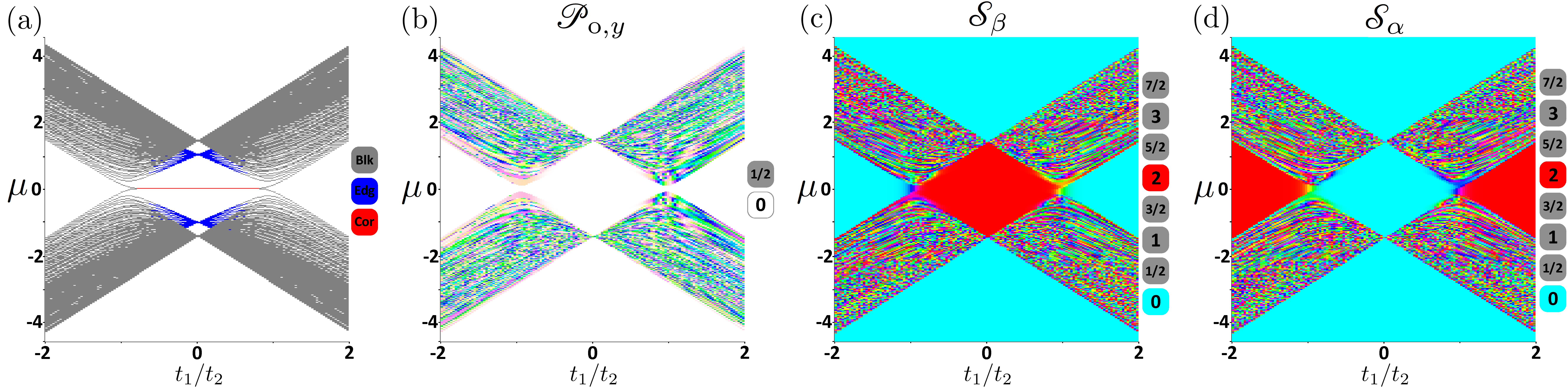}
    \caption{\textbf{(a)} Energy levels of the QI model as a function of $t_1/t_2$. Grey colored states are bulk states; blue colored states are localized at the edge, and the red colored states are localized at the corners.    The edge states vanishes on a torus, and the corner states vanishes on either a torus or a cylinder. 
    \textbf{(b)} Numerically calculated $\mathscr{P}_{\OO,y}$, on the edge of a cylinder following Eq.~\eqref{eq:correctedgeC} for $L_y=15,16$. We set up the radius of $W$ such that  $n_{W,\alpha}=6L_y$, and $n_{W,\beta}=6.5L_y$. $\OO=\alpha,\beta$ extract the same $\mathscr{P}_{\OO,y}$.
    \textbf{(c)(d)} Numerically calculated $\SO$ following Eq.~\eqref{eq:qiCorner}. $W$ is defined similarly to that of Fig.~\ref{fig:corner_charge_num}(b) on a lattice where $\vec{L}_{\OO}\simeq[(0,0)]$. Gray color label represent possible values of $\SO$ and $\PO$ but never appeared in the spectrum. Note that $\mathscr{S}_{\OO}$ and $\vec{\mathscr{P}}_{\OO}$ give quantized values only when the bulk is gapped, and regardless of whether the boundary is gapped or not.
 }
    \label{fig:qi_data}
\end{figure*}

In light of our theory of corner charges, it is now appropriate to revisit earlier studies that have investigated corner charges in higher order topological insulators (HOTIs) and quadrupole insulators (QIs) \cite{wladimir2017quantized,schindler2018higher,Benalcazar2019HOTI,roy2020dislocation,maymann2022prb,hirsbrunner2023crystalline}. In this section we demonstrate how these quantized corner charges can be completely described using our charge response theory in the case where the symmetry under consideration involves $U(1)$ charge conservation, $\mathbb{Z}^2$ translation symmetry, and $\mathbb{Z}_M$ point group rotational symmetry, as in this paper. An important conclusion is that the discrete shift $\mathscr{S}_{\OO}$ and quantized electric polarization $\vec{\mathscr{P}}_\OO$ are the invariants that completely account for the corner charge phenomena; quadrupole and higher multipole moments are not necessary to describe these phenomena. Furthermore, as we will see, our results also highlight how there is no natural notion of a trivial vs. non-trivial quadrupole insulator, since a shift of the origin $\OO$ can relate zero and non-zero values of $\mathscr{S}_{\OO}$ and $\vec{\mathscr{P}}_{\OO}$. 

The first QI model, as defined in \cite{wladimir2017quantized} is a $C_4$ symmetric tight binding model where the bands have Chern number $C=0$. The choice of unit cell  and the hopping parameters are shown in Fig.~\ref{fig:qiUnit}. We proceed to calculate $\mathscr{S}_{\OO}$ and $\vec{\mathscr{P}}_{\OO}$ in this model. 

First, we analyze the band structure. When corners and edges are present in the lattice, localized corner modes and delocalized edge states emerge. These states are highlighted in Fig.~\ref{fig:qi_data}(a). Since the topological invariant $\mathscr{S}_{\OO}$ $\PO$ and $\nu$ are intrinsically bulk invariants, they remain unaltered upon filling these edge and corner states.

To calculate $\PO$ we set up the system on a cylinder similar to the procedure outlined in Sec.~\ref{sec:edgeC}. Since $C=0$ for each band, the edge charge equation \eqref{eq:cylinderC} simplifies to 

\begin{equation}\label{eq:correctedgeC}
    Q_{W} = L_y \mathscr{P}_{\OO,y} + \nu n_{W,\OO}\mod 1
\end{equation}
Again, $\nu n_{W,\alpha}$ is always an integer, and $\nu n_{W,\beta}=\nu n_{W,\alpha}+L_y/2$. A direct numerical calculation of $\mathscr{P}_{\OO,y}$ using Eq.~\eqref{eq:correctedgeC} is shown in Fig.~\ref{fig:qi_data}. We found $\mathscr{P}_{\alpha,y}=\mathscr{P}_{\beta,y}=0\mod 1$ for every bulk gapped phase. Because of the $C_4$ symmetry, we have $\mathscr{P}_{\alpha,x}=\mathscr{P}_{\beta,x}=0\mod 1$ as well.

As a sanity check, \cite{zhang2022pol} derived that under a shift of origin $\OO\rightarrow\OO+\vec{v}$, $\vec{\mathscr{P}}_{\OO}$ should shift following the equation:
\begin{equation}\label{eq:Pshift}
    \vec{\mathscr{P}}_{\OO + \vec{v}} =\vec{\mathscr{P}}_{\OO}  + \kappa (-v_y,v_x),
\end{equation}
where $\kappa = \nu - C \phi/2\pi = \nu$ in this case. The numerically calculated $\vec{\mathscr{P}}_{\OO}$ in this case indeed satisfies Eq. \ref{eq:Pshift}.

Next, we set up the calculation of $\mathscr{S}$ through corner charge in a ribbon geometry similar to Fig.~\ref{fig:corner_charge_num}(b). In this case $\Omega_{\text{cor}}=-\frac{\pi}{2}$, $C=0$, $\vec{\mathscr{P}}_{\OO}=(0,0)$. Eq.~\eqref{eq:cornerC} simplifies to

\begin{equation}\label{eq:qiCorner}
    Q_{W} =  -\frac{1}{4}\mathscr{S}_{\OO}+ \nu n_{W,\OO}\mod 1.
\end{equation}
The ribbon lattice of the QI is constructed by replacing each plaquette in Fig.~\ref{fig:corner_charge_num}(b) by the QI unit cell. 
A direct numerical calculation of $\mathscr{S}_{\OO}$ is shown in Fig.~\ref{fig:qi_data}. 

As a sanity check, the numerical value of $\mathscr{S}_{\alpha}$ and $\mathscr{S}_{\beta}$ satisfy their proven relation \cite{zhang2022pol}

\begin{align}
\mathscr{S}_{\beta}=\mathscr{S}_{\alpha}+4\mathscr{P}_{\alpha,y}-\kappa.
\end{align}

We have thereby shown that the corner charge in a HOTI model can be described using our theory of the charge response.

In the HOTI literature, insulators with non-zero corner (or disclination) charges are classified as topologically non-trivial HOTIs, while those with zero corner (or disclination) charges are deemed trivial. However, this binary classification is unnatural. For instance, in the square-lattice Hofstadter model at full filling, $\mathscr{S}_{\alpha}=0\mod 4$ but  $\mathscr{S}_{\beta}=1\mod 4$ \cite{zhang2022pol}. This means that a plaquette-centered corner would contribute zero charge, while a vertex-centered corner would contribute a charge of $\pm 1/8 \mod 1$. Therefore, whether a HOTI is deemed trivial or non-trivial depends on what high symmetry point $\OO$ is chosen for the corner and disclination cores. 
Furthermore, one can show that even the simplest model, the one-band square lattice tight-binding model has $\mathscr{S}_{\beta}=1\mod 4$ and $\mathscr{S}_{\alpha} = 0 \mod 4$ at full filling. Therefore, we avoid making the binary distinction between trivial and non-trivial insulators.

In previous studies of QIs, quantized $\frac{1}{2}$ corner charge is calculated in the parameter range $|\frac{t_1}{t_2}|<1$. Within the framework of TQFT, this is equivalent to asserting that $\mathscr{S}_{\beta}=2 \mod 4$ within this parameter range as dictated by Eq.~\eqref{eq:qiCorner}. 

\section{Discussion}

In this paper, we have explored how topological invariants, specifically the discrete shift $\SO$ and the electric polarization $\PO$, manifest in the boundary and corner charges of crystalline Chern insulators. Our main result, the full charge response in Eq.~\eqref{eq:charge}, provides a unified framework for understanding the contributions to boundary and corner charges as arising from a combination of the topological invariants $\{C,\nu, \SO,\PO\}$. 

Importantly, our theory is applicable to systems with gapless boundaries such as Chern insulators, where traditional approaches to defining polarization are often problematic. By properly defining the full boundary charge modulo 1, we circumvent these issues and provide a consistent definition of a topologically protected polarization $\PO$ even in the presence of the gapless edge mode.

One of our key insights from our work is that the boundary charge and the charge associated with bulk defects such as disclination and dislocations can be treated on a equal footing. Specifically, we have used the same geometrical measure $\Gamma$ and $\vec{L}_{\OO}$ which applies to both bulk defects and boundary. 

As an application of our result, our charge response naturally describes the corner charges in higher-order topological insulators. We find that the discrete shift $\SO$ fully accounts for the corner charges, contrary to the multipolar moments which is widely used in the literature.

Though in this paper we focus on the square lattice, which has $C_4$ rotational point group symmetry, our results naturally extend to other point group symmetries $C_M$ for $M=2,3,6$. Note that for the $M = 6$ case, the polarization is quantized to a single trivial value \cite{manjunath2021cgt}. 

We can also use our methods to define an electric polarization for Chern insulators in the case of no point group symmetry, $M = 1$. In this case, we can pick any point $\OO$ in the unit cell, and we must not allow corners and disclinations since their response relies on rotational symmetry, so we are limited to geometries without corners. To use our formulas, we then need to define $n_{W,\OO}$ for any point $\OO$ in the unit cell. This can be done by independently computing the polarization in the case of $C = 0$ using the localized Wannier functions, and then fitting to the formula for $Q_W$ to obtain $n_{W,\OO}$. However this procedure requires a notion of distance between the origin $\OO$ and the Wannier orbitals, and this information does not come directly from the Hamiltonian, but rather needs to come from additional input about the system being described. 

We close by pointing out interesting future directions. First, it is important to understand the relationship between the many-body electirc polarization defined here for Chern insulators and the one defined in \cite{coh2009}, which required an arbitrary choice of origin in the Brillouin Zone and used the single-particle Berry phase theory of polarization. Furthermore, it is important to further study $\SO$ and $\PO$ in the fractional Chern insulators. These were studied using topological field theory methods in Refs. \cite{manjunath2021cgt,manjunath2023classif}. Given the anyon data and the $M$ fold point group symmetry, these $\SO$ and $\PO$ could further fractionalize. These symmetry fractionalization data are encapsulated in the spin vector $s$ and the discrete torsion vector $\vec{t}$. 
They also manifest as the charge response of the bulk defects and boundaries and could in principle be extracted in microscopic models using similar methods. 

\acknowledgments

We thank Naren Manjunath for discussions, comments on the draft, and collaboration on related work. This work is supported by NSF DMR-2345644 and by NSF QLCI grant OMA-2120757 through the Institute for Robust Quantum Simulation (RQS).

\appendix

\section{$C=0$ calculation of the unit cell measure $\vec{n}_{\OO}$, $m_{\OO}$}\label{sec:zeroC}

\begin{figure}[t]
    \centering
    \includegraphics[width=8cm]{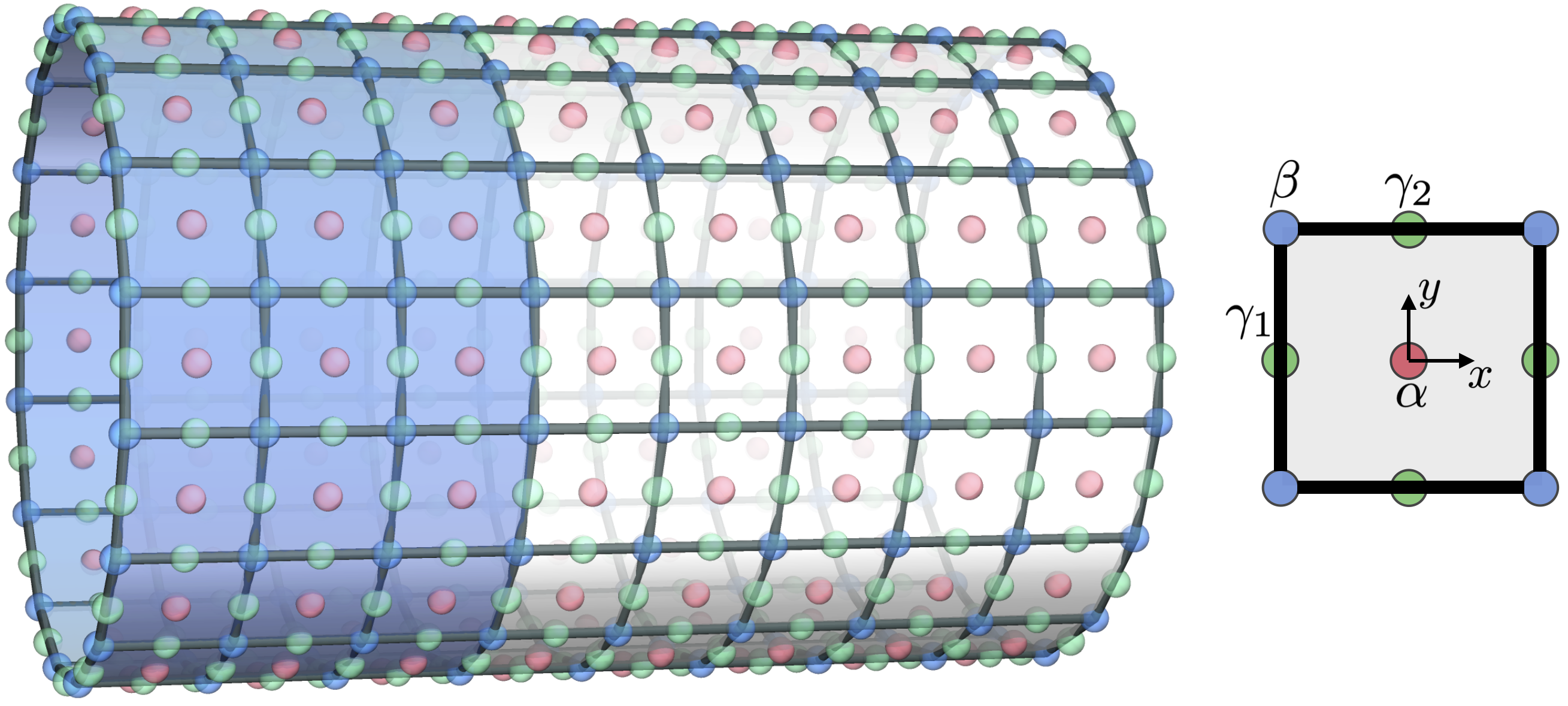}
    \caption{Square lattice on a cylinder with MWPs $\{\alpha,\beta,\gamma_1,\gamma_2\}$ labeled explicitly.
    }
    \label{fig:cylinderapp}
\end{figure}

When $C=0$ the system can be adiabatically connected to an atomic insulator in which the electron wave function consists of localized Wannier orbitals placed at the maximal Wyckoff positions. $\PO$, $\SO$ can be analytically calculated in such a state in terms of the classical charge distribution. In this section we use this fact to calculate $\vec{n}_{\OO}$ and $m_{\OO}$ by fitting to the charge response equation.

In the discussion below we only discuss the $C_4$ symmetric unit cell but the whole calculation can be straightforwardly generalized to $C_2,C_3,C_6$ symmetric unit cells. We denote the positive integers $N_{\OO}$ as the number of filled orbitals at the MWPs $\OO\in\{\alpha,\beta,\gamma\}$. As derived in \cite{zhang2022pol}, $\SO$ for the $C_4$ symmetric MWPs $\{\alpha,\beta\}$ can be expressed as

\begin{align}
    \mathscr{S}_{\alpha}&=N_{\alpha} \mod 4,\nonumber\\
    \mathscr{S}_{\beta}&=N_{\beta} \mod 4,
\end{align}

and $\PO$ is expressed as, modulo 1,
\begin{align}\label{eq:pc0}
    \vec{\mathscr{P}}_{\alpha}&=\frac{N_{\beta}+N_{\gamma}}{2}(1,1)\nonumber \\
    \vec{\mathscr{P}}_{\beta}&=\frac{N_{\alpha}+N_{\gamma}}{2}(1,1)
\end{align}

We first calculate $\vec{n}_{\OO}=(n_{\OO,x},n_{\OO,y})$. Consider a $C=0$ model defined on a cylinder shown in Fig.~\ref{fig:cylinderapp}. The charge response is of the form

\begin{equation}
    Q_{W} = \vec{L}_{\OO}\cdot\vec{\mathscr{P}}_{\OO} + \nu (k+\vec{L}_{\OO}\cdot\vec{n}_{\OO}) \mod 1,
\end{equation}
Plugging in Eq.~\eqref{eq:pc0}, and $\nu=N_{\alpha}+N_{\beta}+2N_{\gamma}$, $k=3L_y$, we have

\begin{align}\label{eq:q_in_N}
    Q_{W} &= L_y [\frac{N_{\beta}+N_{\gamma}}{2} + (N_{\alpha}+N_{\beta}+2N_{\gamma})(3+ n_{\alpha,y})]\mod 1\nonumber\\
    &= L_y [\frac{N_{\alpha}+N_{\gamma}}{2} + (N_{\alpha}+N_{\beta}+2N_{\gamma})(3+n_{\beta,y})]\mod 1
\end{align}

On the other hand, by explicit counting of the orbitals in Fig.~\ref{fig:cylinderapp},
\begin{equation}
    Q_W=L_y[3N_{\alpha}+(3+1/2)N_{\beta}+(3+3+1/2)N_{\gamma}]\mod 1
\end{equation}

Plugging in Eq.~\eqref{eq:q_in_N}, we can solve for $n_{\OO,y}$. For generic $L_y\in \Z$ 

\begin{align}
    n_{\alpha,y}&=0 \mod 1\nonumber\\
    n_{\beta,y}&=1/2 \mod 1,
\end{align}
and $n_{\OO,x}=n_{\OO,y}$ for $\OO=\{\alpha,\beta\}$ because of the $C_4$ symmetry.

Note that in order to define $\delta\Phi_{W,\OO}$, we need to know $\vec{n}_{\OO}$ absolutely instead of modulo 1. To resolve this, we can simply pick an arbitrary lift to $\mathbb{R}^2$. Our choice is

\begin{align}
    \vec{n}_{\alpha}&=(0,0)\nonumber\\
    \vec{n}_{\beta}&=(1/2,1/2). 
\end{align}
This recovers the $\vec{n}_{\OO}$ entry in Tab.~\ref{table:nuc}. We could pick a different lift: $\vec{n}_{\OO}\rightarrow \vec{n}_{\OO}+\vec{v}$ for an integer vector $\vec{v}$. Under this choice, 
$n_{W,\OO} \rightarrow n_{W,\OO} + \vec{v}\cdot \vec{L}_{\OO}$ and $\delta \Phi_{W,\OO}\rightarrow \delta \Phi_{W,\OO}-\phi\vec{v}\cdot \vec{L}_{\OO}$ so that $\nu n_{W,\OO} + \frac{C}{2\pi} \delta \Phi_{W,\OO} \rightarrow\nu n_{W,\OO} + \frac{C}{2\pi} \delta \Phi_{W,\OO}+\vec{v}\cdot \vec{L}_{\OO}\kappa$, where the shift $\vec{v}\cdot \vec{L}_{\OO}\kappa$ is an integer. Therefore, the invariants extracted using Eq.~\eqref{eq:charge} are unchanged under this change of lift. 

Next, we calculate $m_{\OO}$. One could perform a similar calculation to that of 
$\vec{n}_{\OO}$ as above, and obtain
$m_{\alpha}=1$ and $m_{\beta}=0$. Here, we provide a more geometrically illuminating solution. 

\begin{figure}[t]
    \centering
    \includegraphics[width=7.5cm]{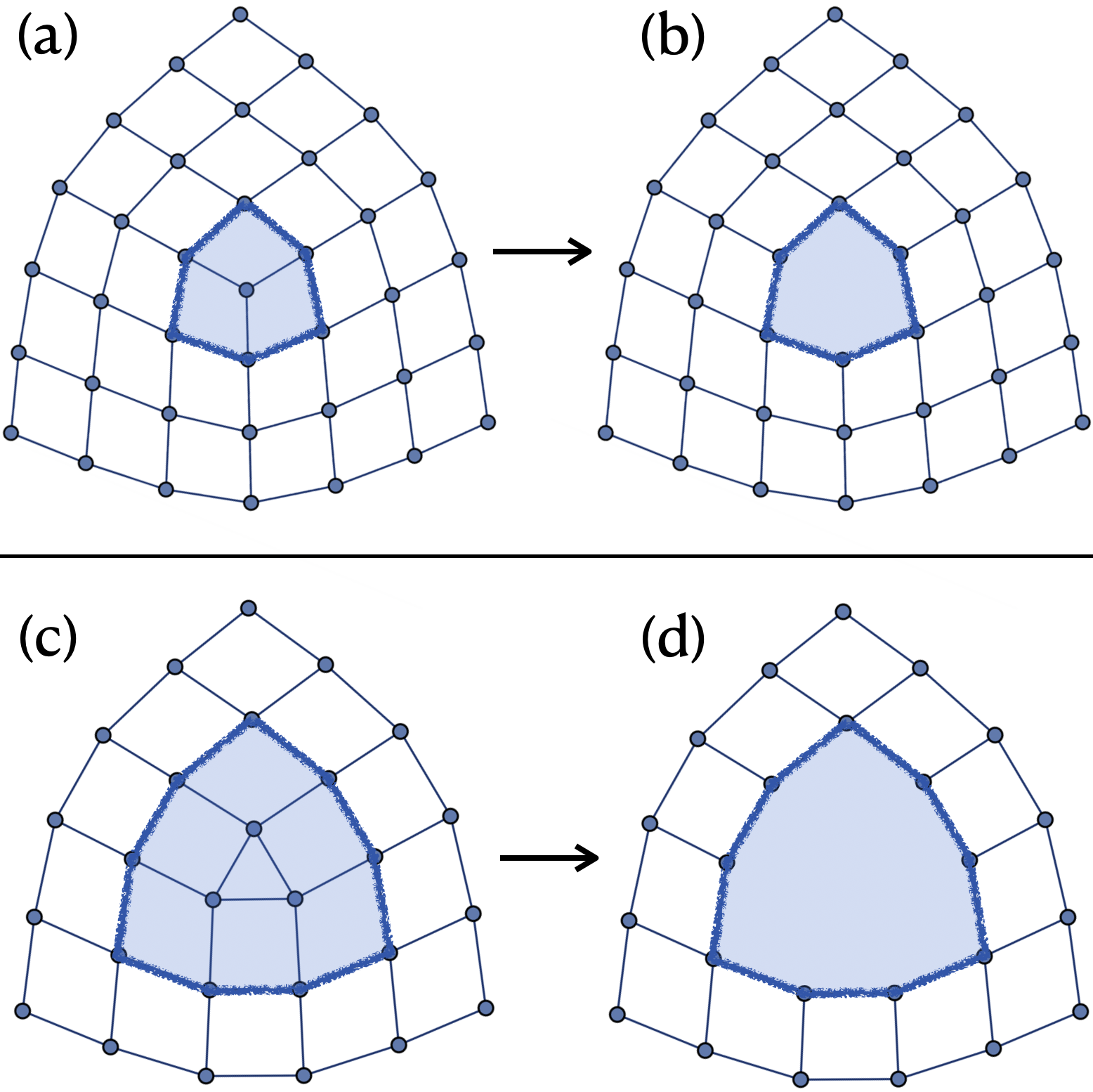}
    \caption{\textbf{(a)(b)} Inserting a trivial defect by removing a site at the core of a pure $\OO=\beta$ disclination. $n_{W,\beta}=3$ before and after inserting the trivial defect. \textbf{(c)(d)} Inserting a trivial defect by removing three sites at the core of a pure $\OO=\alpha$ disclination. $n_{W,\beta}=6\frac{3}{4}$ before and after inserting the trivial defect.
    }
    \label{fig:dtc}
\end{figure}

In Sec.~\ref{sec:general}, we have argued that a ribbon with $\Omega_{\text{cor}}$ total corner angle can be seen as a disclination with $2\pi +\Omega_{\text{cor}}$ disclination angle, and vice versa. Consider two $\Omega=\frac{\pi}{2}$ pure disclinations with holes in the middle shown in Fig.~\ref{fig:dtc} for $\OO=\{\alpha,\beta\}$. They are created by removing sites from the disclination core which amounts to inserting trivial defects. During this process, the number of unit cells within the region $W$ defined in the figure does not change. We treat Fig.~\ref{fig:dtc} (b),(d) as having three corners with corner angle $-\pi/2$ each, and total corner angle $\Omega_{\text{cor}}=-\frac{3\pi}{2}$. Now we solve for $m_{\OO}$ by matching the fractional part of $n_{W,\OO}$.

\begin{align}
    \vec{L}_{\beta}\cdot\vec{n}_{\beta}+\frac{3}{4}m_{\beta}=3 \mod 1\nonumber\\
    \vec{L}_{\alpha}\cdot\vec{n}_{\alpha}+\frac{3}{4}m_{\alpha}=6\frac{3}{4} \mod 1.
\end{align}

Since both disclinations are pure with trivial Burgers vector, $\vec{L}_{\OO}$ is also in the trivial class, and therefore $\vec{n}_{\OO}$ does not contribute to fractional part of $n_{W,\OO}$. Therefore, we have 

\begin{align}\label{eq:ncor}
    m_{\beta}=0 \mod 4/3\nonumber\\
    m_{\alpha}=1 \mod 4/3
\end{align}

Eq.~\eqref{eq:ncor} determines $m_{\OO}$ modulo $4/3$. Now consider a trivial defect on a clean lattice shown in Fig.~\ref{fig:trivial} where the hole consists of 4 corners with total corner angle $-2\pi$ and no fractional unit cell, therefore we have

\begin{align}\label{eq:ncor2}
    &m_{\beta}=0 \mod 1\nonumber\\
    &m_{\alpha}=0 \mod 1,
\end{align}

Together with Eq.~\eqref{eq:ncor}, we are able to determine $m_{\beta}$ modulo $lcm(\frac{4}{3},1)=4$, where $lcm$ is the lowest common multiple:

\begin{align}
    &m_{\beta}=0 \mod 4\nonumber\\
    &m_{\alpha}=1 \mod 4.
\end{align}

Again, in order to define $\delta{\Phi}_{W,\OO}$, we need to know $m_{\OO}$ absolutely, and this requires picking a lift to $\Z$, our choice is

\begin{align}
   &m_{\beta}=0 \nonumber\\
   &m_{\alpha}=1.
\end{align}
This reproduces the $m_{\OO}$ entry in Tab.~\ref{table:nuc}. Similar to $\vec{n}_{\OO}$, upon picking a different lift $m_{\OO}\rightarrow m_{\OO}+4r$ for a integer $r$.  Under this choice, 
$n_{W,\OO} \rightarrow n_{W,\OO} + r \frac{\Gamma+2\pi}{2\pi}$ and $\delta \Phi_{W,\OO}\rightarrow \delta \Phi_{W,\OO}-\phi r \frac{\Gamma+2\pi}{2\pi}$ so that $\nu n_{W,\OO} + \frac{C}{2\pi} \delta \Phi_{W,\OO} \rightarrow\nu n_{W,\OO} + \frac{C}{2\pi} \delta \Phi_{W,\OO}+ r \frac{\Gamma+2\pi}{2\pi}\kappa$, where the shift $ r \frac{\Gamma+2\pi}{2\pi}\kappa$ is an integer.
Therefore, the invariants extracted using Eq.~\eqref{eq:charge} are unchanged under this change of lift. 

\bibliography{bibliography}

\begin{thebibliography}{65}%
\makeatletter
\providecommand \@ifxundefined [1]{%
 \@ifx{#1\undefined}
}%
\providecommand \@ifnum [1]{%
 \ifnum #1\expandafter \@firstoftwo
 \else \expandafter \@secondoftwo
 \fi
}%
\providecommand \@ifx [1]{%
 \ifx #1\expandafter \@firstoftwo
 \else \expandafter \@secondoftwo
 \fi
}%
\providecommand \natexlab [1]{#1}%
\providecommand \enquote  [1]{``#1''}%
\providecommand \bibnamefont  [1]{#1}%
\providecommand \bibfnamefont [1]{#1}%
\providecommand \citenamefont [1]{#1}%
\providecommand \href@noop [0]{\@secondoftwo}%
\providecommand \href [0]{\begingroup \@sanitize@url \@href}%
\providecommand \@href[1]{\@@startlink{#1}\@@href}%
\providecommand \@@href[1]{\endgroup#1\@@endlink}%
\providecommand \@sanitize@url [0]{\catcode `\\12\catcode `\$12\catcode
  `\&12\catcode `\#12\catcode `\^12\catcode `\_12\catcode `\%12\relax}%
\providecommand \@@startlink[1]{}%
\providecommand \@@endlink[0]{}%
\providecommand \url  [0]{\begingroup\@sanitize@url \@url }%
\providecommand \@url [1]{\endgroup\@href {#1}{\urlprefix }}%
\providecommand \urlprefix  [0]{URL }%
\providecommand \Eprint [0]{\href }%
\providecommand \doibase [0]{http://dx.doi.org/}%
\providecommand \selectlanguage [0]{\@gobble}%
\providecommand \bibinfo  [0]{\@secondoftwo}%
\providecommand \bibfield  [0]{\@secondoftwo}%
\providecommand \translation [1]{[#1]}%
\providecommand \BibitemOpen [0]{}%
\providecommand \bibitemStop [0]{}%
\providecommand \bibitemNoStop [0]{.\EOS\space}%
\providecommand \EOS [0]{\spacefactor3000\relax}%
\providecommand \BibitemShut  [1]{\csname bibitem#1\endcsname}%
\let\auto@bib@innerbib\@empty
%</preamble>
\bibitem [{\citenamefont {Wen}(2002)}]{wen2002quantum}%
  \BibitemOpen
  \bibfield  {author} {\bibinfo {author} {\bibfnamefont {X.-G.}\ \bibnamefont
  {Wen}},\ }\href@noop {} {\bibfield  {journal} {\bibinfo  {journal} {Physical
  Review B}\ }\textbf {\bibinfo {volume} {65}},\ \bibinfo {pages} {165113}
  (\bibinfo {year} {2002})}\BibitemShut {NoStop}%
\bibitem [{\citenamefont {Wen}(2004)}]{wen04}%
  \BibitemOpen
  \bibfield  {author} {\bibinfo {author} {\bibfnamefont {X.-G.}\ \bibnamefont
  {Wen}},\ }\href@noop {} {\emph {\bibinfo {title} {Quantum Field Theory of
  Many-Body Systems}}}\ (\bibinfo  {publisher} {Oxford Univ. Press},\ \bibinfo
  {address} {Oxford},\ \bibinfo {year} {2004})\BibitemShut {NoStop}%
\bibitem [{\citenamefont {Hasan}\ and\ \citenamefont {Kane}(2010)}]{hasan2010}%
  \BibitemOpen
  \bibfield  {author} {\bibinfo {author} {\bibfnamefont {M.~Z.}\ \bibnamefont
  {Hasan}}\ and\ \bibinfo {author} {\bibfnamefont {C.~L.}\ \bibnamefont
  {Kane}},\ }\href {\doibase 10.1103/RevModPhys.82.3045} {\bibfield  {journal}
  {\bibinfo  {journal} {Rev. Mod. Phys.}\ }\textbf {\bibinfo {volume} {82}},\
  \bibinfo {pages} {3045} (\bibinfo {year} {2010})}\BibitemShut {NoStop}%
\bibitem [{\citenamefont {Fu}(2011)}]{fu2011topological}%
  \BibitemOpen
  \bibfield  {author} {\bibinfo {author} {\bibfnamefont {L.}~\bibnamefont
  {Fu}},\ }\href@noop {} {\bibfield  {journal} {\bibinfo  {journal} {Physical
  review letters}\ }\textbf {\bibinfo {volume} {106}},\ \bibinfo {pages}
  {106802} (\bibinfo {year} {2011})}\BibitemShut {NoStop}%
\bibitem [{\citenamefont {Barkeshli}\ and\ \citenamefont
  {Qi}(2012)}]{barkeshli2012a}%
  \BibitemOpen
  \bibfield  {author} {\bibinfo {author} {\bibfnamefont {M.}~\bibnamefont
  {Barkeshli}}\ and\ \bibinfo {author} {\bibfnamefont {X.-L.}\ \bibnamefont
  {Qi}},\ }\href {\doibase 10.1103/PhysRevX.2.031013} {\bibfield  {journal}
  {\bibinfo  {journal} {Phys. Rev. X}\ }\textbf {\bibinfo {volume} {2}},\
  \bibinfo {pages} {031013} (\bibinfo {year} {2012})},\ \Eprint
  {http://arxiv.org/abs/arXiv:1112.3311} {arXiv:1112.3311} \BibitemShut
  {NoStop}%
\bibitem [{\citenamefont {Essin}\ and\ \citenamefont
  {Hermele}(2013)}]{Essin2013SF}%
  \BibitemOpen
  \bibfield  {author} {\bibinfo {author} {\bibfnamefont {A.~M.}\ \bibnamefont
  {Essin}}\ and\ \bibinfo {author} {\bibfnamefont {M.}~\bibnamefont
  {Hermele}},\ }\href {\doibase 10.1103/PhysRevB.87.104406} {\bibfield
  {journal} {\bibinfo  {journal} {Phys. Rev. B}\ }\textbf {\bibinfo {volume}
  {87}},\ \bibinfo {pages} {104406} (\bibinfo {year} {2013})}\BibitemShut
  {NoStop}%
\bibitem [{\citenamefont {Essin}\ and\ \citenamefont
  {Hermele}(2014)}]{Essin2014spect}%
  \BibitemOpen
  \bibfield  {author} {\bibinfo {author} {\bibfnamefont {A.~M.}\ \bibnamefont
  {Essin}}\ and\ \bibinfo {author} {\bibfnamefont {M.}~\bibnamefont
  {Hermele}},\ }\href {\doibase 10.1103/PhysRevB.90.121102} {\bibfield
  {journal} {\bibinfo  {journal} {Phys. Rev. B}\ }\textbf {\bibinfo {volume}
  {90}},\ \bibinfo {pages} {121102} (\bibinfo {year} {2014})}\BibitemShut
  {NoStop}%
\bibitem [{\citenamefont {Barkeshli}\ \emph
  {et~al.}(2019{\natexlab{a}})\citenamefont {Barkeshli}, \citenamefont
  {Bonderson}, \citenamefont {Cheng},\ and\ \citenamefont
  {Wang}}]{barkeshli2019}%
  \BibitemOpen
  \bibfield  {author} {\bibinfo {author} {\bibfnamefont {M.}~\bibnamefont
  {Barkeshli}}, \bibinfo {author} {\bibfnamefont {P.}~\bibnamefont
  {Bonderson}}, \bibinfo {author} {\bibfnamefont {M.}~\bibnamefont {Cheng}}, \
  and\ \bibinfo {author} {\bibfnamefont {Z.}~\bibnamefont {Wang}},\ }\href
  {\doibase 10.1103/PhysRevB.100.115147} {\bibfield  {journal} {\bibinfo
  {journal} {Phys. Rev. B}\ }\textbf {\bibinfo {volume} {100}},\ \bibinfo
  {pages} {115147} (\bibinfo {year} {2019}{\natexlab{a}})},\ \Eprint
  {http://arxiv.org/abs/arXiv:1410.4540} {arXiv:1410.4540} \BibitemShut
  {NoStop}%
\bibitem [{\citenamefont {Benalcazar}\ \emph {et~al.}(2014)\citenamefont
  {Benalcazar}, \citenamefont {Teo},\ and\ \citenamefont
  {Hughes}}]{Benalcazar2014}%
  \BibitemOpen
  \bibfield  {author} {\bibinfo {author} {\bibfnamefont {W.~A.}\ \bibnamefont
  {Benalcazar}}, \bibinfo {author} {\bibfnamefont {J.~C.~Y.}\ \bibnamefont
  {Teo}}, \ and\ \bibinfo {author} {\bibfnamefont {T.~L.}\ \bibnamefont
  {Hughes}},\ }\href {\doibase 10.1103/PhysRevB.89.224503} {\bibfield
  {journal} {\bibinfo  {journal} {Phys. Rev. B}\ }\textbf {\bibinfo {volume}
  {89}},\ \bibinfo {pages} {224503} (\bibinfo {year} {2014})}\BibitemShut
  {NoStop}%
\bibitem [{\citenamefont {Ando}\ and\ \citenamefont {Fu}(2015)}]{ando2015}%
  \BibitemOpen
  \bibfield  {author} {\bibinfo {author} {\bibfnamefont {Y.}~\bibnamefont
  {Ando}}\ and\ \bibinfo {author} {\bibfnamefont {L.}~\bibnamefont {Fu}},\
  }\href@noop {} {\bibfield  {journal} {\bibinfo  {journal} {Annu. Rev.
  Condens. Matter Phys.}\ }\textbf {\bibinfo {volume} {6}},\ \bibinfo {pages}
  {361} (\bibinfo {year} {2015})}\BibitemShut {NoStop}%
\bibitem [{\citenamefont {Qi}\ and\ \citenamefont {Fu}(2015)}]{YangPRL2015}%
  \BibitemOpen
  \bibfield  {author} {\bibinfo {author} {\bibfnamefont {Y.}~\bibnamefont
  {Qi}}\ and\ \bibinfo {author} {\bibfnamefont {L.}~\bibnamefont {Fu}},\ }\href
  {\doibase 10.1103/PhysRevLett.115.236801} {\bibfield  {journal} {\bibinfo
  {journal} {Phys. Rev. Lett.}\ }\textbf {\bibinfo {volume} {115}},\ \bibinfo
  {pages} {236801} (\bibinfo {year} {2015})}\BibitemShut {NoStop}%
\bibitem [{\citenamefont {Watanabe}\ \emph {et~al.}(2015)\citenamefont
  {Watanabe}, \citenamefont {Po}, \citenamefont {Vishwanath},\ and\
  \citenamefont {Zaletel}}]{watanabe2015filling}%
  \BibitemOpen
  \bibfield  {author} {\bibinfo {author} {\bibfnamefont {H.}~\bibnamefont
  {Watanabe}}, \bibinfo {author} {\bibfnamefont {H.~C.}\ \bibnamefont {Po}},
  \bibinfo {author} {\bibfnamefont {A.}~\bibnamefont {Vishwanath}}, \ and\
  \bibinfo {author} {\bibfnamefont {M.}~\bibnamefont {Zaletel}},\ }\href@noop
  {} {\bibfield  {journal} {\bibinfo  {journal} {Proceedings of the National
  Academy of Sciences}\ }\textbf {\bibinfo {volume} {112}},\ \bibinfo {pages}
  {14551} (\bibinfo {year} {2015})}\BibitemShut {NoStop}%
\bibitem [{\citenamefont {Watanabe}\ \emph {et~al.}(2016)\citenamefont
  {Watanabe}, \citenamefont {Po}, \citenamefont {Zaletel},\ and\ \citenamefont
  {Vishwanath}}]{watanabe2016filling}%
  \BibitemOpen
  \bibfield  {author} {\bibinfo {author} {\bibfnamefont {H.}~\bibnamefont
  {Watanabe}}, \bibinfo {author} {\bibfnamefont {H.~C.}\ \bibnamefont {Po}},
  \bibinfo {author} {\bibfnamefont {M.~P.}\ \bibnamefont {Zaletel}}, \ and\
  \bibinfo {author} {\bibfnamefont {A.}~\bibnamefont {Vishwanath}},\
  }\href@noop {} {\bibfield  {journal} {\bibinfo  {journal} {Physical review
  letters}\ }\textbf {\bibinfo {volume} {117}},\ \bibinfo {pages} {096404}
  (\bibinfo {year} {2016})}\BibitemShut {NoStop}%
\bibitem [{\citenamefont {Chiu}\ \emph {et~al.}(2016)\citenamefont {Chiu},
  \citenamefont {Teo}, \citenamefont {Schnyder},\ and\ \citenamefont
  {Ryu}}]{Chiu2016review}%
  \BibitemOpen
  \bibfield  {author} {\bibinfo {author} {\bibfnamefont {C.-K.}\ \bibnamefont
  {Chiu}}, \bibinfo {author} {\bibfnamefont {J.~C.~Y.}\ \bibnamefont {Teo}},
  \bibinfo {author} {\bibfnamefont {A.~P.}\ \bibnamefont {Schnyder}}, \ and\
  \bibinfo {author} {\bibfnamefont {S.}~\bibnamefont {Ryu}},\ }\href {\doibase
  10.1103/RevModPhys.88.035005} {\bibfield  {journal} {\bibinfo  {journal}
  {Rev. Mod. Phys.}\ }\textbf {\bibinfo {volume} {88}},\ \bibinfo {pages}
  {035005} (\bibinfo {year} {2016})}\BibitemShut {NoStop}%
\bibitem [{\citenamefont {Hermele}\ and\ \citenamefont
  {Chen}(2016)}]{hermele2016}%
  \BibitemOpen
  \bibfield  {author} {\bibinfo {author} {\bibfnamefont {M.}~\bibnamefont
  {Hermele}}\ and\ \bibinfo {author} {\bibfnamefont {X.}~\bibnamefont {Chen}},\
  }\href {\doibase 10.1103/PhysRevX.6.041006} {\bibfield  {journal} {\bibinfo
  {journal} {Phys. Rev. X}\ }\textbf {\bibinfo {volume} {6}},\ \bibinfo {pages}
  {041006} (\bibinfo {year} {2016})}\BibitemShut {NoStop}%
\bibitem [{\citenamefont {Barkeshli}\ \emph
  {et~al.}(2019{\natexlab{b}})\citenamefont {Barkeshli}, \citenamefont
  {Bonderson}, \citenamefont {Cheng}, \citenamefont {Jian},\ and\ \citenamefont
  {Walker}}]{barkeshli2019tr}%
  \BibitemOpen
  \bibfield  {author} {\bibinfo {author} {\bibfnamefont {M.}~\bibnamefont
  {Barkeshli}}, \bibinfo {author} {\bibfnamefont {P.}~\bibnamefont
  {Bonderson}}, \bibinfo {author} {\bibfnamefont {M.}~\bibnamefont {Cheng}},
  \bibinfo {author} {\bibfnamefont {C.-M.}\ \bibnamefont {Jian}}, \ and\
  \bibinfo {author} {\bibfnamefont {K.}~\bibnamefont {Walker}},\ }\href
  {\doibase 10.1007/s00220-019-03475-8} {\bibfield  {journal} {\bibinfo
  {journal} {Communications in Mathematical Physics}\ } (\bibinfo {year}
  {2019}{\natexlab{b}}),\ 10.1007/s00220-019-03475-8},\ \Eprint
  {http://arxiv.org/abs/arXiv:1612.07792} {arXiv:1612.07792} \BibitemShut
  {NoStop}%
\bibitem [{\citenamefont {Zaletel}\ \emph {et~al.}(2017)\citenamefont
  {Zaletel}, \citenamefont {Lu},\ and\ \citenamefont
  {Vishwanath}}]{zaletel2017}%
  \BibitemOpen
  \bibfield  {author} {\bibinfo {author} {\bibfnamefont {M.~P.}\ \bibnamefont
  {Zaletel}}, \bibinfo {author} {\bibfnamefont {Y.-M.}\ \bibnamefont {Lu}}, \
  and\ \bibinfo {author} {\bibfnamefont {A.}~\bibnamefont {Vishwanath}},\
  }\href {\doibase 10.1103/PhysRevB.96.195164} {\bibfield  {journal} {\bibinfo
  {journal} {Phys. Rev. B}\ }\textbf {\bibinfo {volume} {96}},\ \bibinfo
  {pages} {195164} (\bibinfo {year} {2017})}\BibitemShut {NoStop}%
\bibitem [{\citenamefont {Po}\ \emph {et~al.}(2017)\citenamefont {Po},
  \citenamefont {Vishwanath},\ and\ \citenamefont {Watanabe}}]{Po2017symmind}%
  \BibitemOpen
  \bibfield  {author} {\bibinfo {author} {\bibfnamefont {H.~C.}\ \bibnamefont
  {Po}}, \bibinfo {author} {\bibfnamefont {A.}~\bibnamefont {Vishwanath}}, \
  and\ \bibinfo {author} {\bibfnamefont {H.}~\bibnamefont {Watanabe}},\ }\href
  {\doibase 10.1038/s41467-017-00133-2} {\bibfield  {journal} {\bibinfo
  {journal} {Nature Communications}\ }\textbf {\bibinfo {volume} {8}} (\bibinfo
  {year} {2017}),\ 10.1038/s41467-017-00133-2}\BibitemShut {NoStop}%
\bibitem [{\citenamefont {Song}\ \emph {et~al.}(2017)\citenamefont {Song},
  \citenamefont {Huang}, \citenamefont {Fu},\ and\ \citenamefont
  {Hermele}}]{song2017}%
  \BibitemOpen
  \bibfield  {author} {\bibinfo {author} {\bibfnamefont {H.}~\bibnamefont
  {Song}}, \bibinfo {author} {\bibfnamefont {S.-J.}\ \bibnamefont {Huang}},
  \bibinfo {author} {\bibfnamefont {L.}~\bibnamefont {Fu}}, \ and\ \bibinfo
  {author} {\bibfnamefont {M.}~\bibnamefont {Hermele}},\ }\href {\doibase
  10.1103/PhysRevX.7.011020} {\bibfield  {journal} {\bibinfo  {journal} {Phys.
  Rev. X}\ }\textbf {\bibinfo {volume} {7}},\ \bibinfo {pages} {011020}
  (\bibinfo {year} {2017})}\BibitemShut {NoStop}%
\bibitem [{\citenamefont {Huang}\ \emph {et~al.}(2017)\citenamefont {Huang},
  \citenamefont {Song}, \citenamefont {Huang},\ and\ \citenamefont
  {Hermele}}]{Huang2017}%
  \BibitemOpen
  \bibfield  {author} {\bibinfo {author} {\bibfnamefont {S.-J.}\ \bibnamefont
  {Huang}}, \bibinfo {author} {\bibfnamefont {H.}~\bibnamefont {Song}},
  \bibinfo {author} {\bibfnamefont {Y.-P.}\ \bibnamefont {Huang}}, \ and\
  \bibinfo {author} {\bibfnamefont {M.}~\bibnamefont {Hermele}},\ }\href
  {\doibase 10.1103/PhysRevB.96.205106} {\bibfield  {journal} {\bibinfo
  {journal} {Phys. Rev. B}\ }\textbf {\bibinfo {volume} {96}},\ \bibinfo
  {pages} {205106} (\bibinfo {year} {2017})}\BibitemShut {NoStop}%
\bibitem [{\citenamefont {Shiozaki}\ \emph {et~al.}(2017)\citenamefont
  {Shiozaki}, \citenamefont {Shapourian},\ and\ \citenamefont
  {Ryu}}]{Shiozaki2017point}%
  \BibitemOpen
  \bibfield  {author} {\bibinfo {author} {\bibfnamefont {K.}~\bibnamefont
  {Shiozaki}}, \bibinfo {author} {\bibfnamefont {H.}~\bibnamefont
  {Shapourian}}, \ and\ \bibinfo {author} {\bibfnamefont {S.}~\bibnamefont
  {Ryu}},\ }\href {\doibase 10.1103/physrevb.95.205139} {\bibfield  {journal}
  {\bibinfo  {journal} {Physical Review B}\ }\textbf {\bibinfo {volume} {95}}
  (\bibinfo {year} {2017}),\ 10.1103/physrevb.95.205139},\ \Eprint
  {http://arxiv.org/abs/1609.05970} {arXiv:1609.05970 [cond-mat.str-el]}
  \BibitemShut {NoStop}%
\bibitem [{\citenamefont {Kruthoff}\ \emph {et~al.}(2017)\citenamefont
  {Kruthoff}, \citenamefont {de~Boer}, \citenamefont {van Wezel}, \citenamefont
  {Kane},\ and\ \citenamefont {Slager}}]{Kruthoff2017TCI}%
  \BibitemOpen
  \bibfield  {author} {\bibinfo {author} {\bibfnamefont {J.}~\bibnamefont
  {Kruthoff}}, \bibinfo {author} {\bibfnamefont {J.}~\bibnamefont {de~Boer}},
  \bibinfo {author} {\bibfnamefont {J.}~\bibnamefont {van Wezel}}, \bibinfo
  {author} {\bibfnamefont {C.~L.}\ \bibnamefont {Kane}}, \ and\ \bibinfo
  {author} {\bibfnamefont {R.-J.}\ \bibnamefont {Slager}},\ }\href {\doibase
  10.1103/PhysRevX.7.041069} {\bibfield  {journal} {\bibinfo  {journal} {Phys.
  Rev. X}\ }\textbf {\bibinfo {volume} {7}},\ \bibinfo {pages} {041069}
  (\bibinfo {year} {2017})}\BibitemShut {NoStop}%
\bibitem [{\citenamefont {Bradlyn}\ \emph {et~al.}(2017)\citenamefont
  {Bradlyn}, \citenamefont {Elcoro},\ and\ \citenamefont
  {Jennifer~Cano}}]{Bradlyn2017tqc}%
  \BibitemOpen
  \bibfield  {author} {\bibinfo {author} {\bibfnamefont {B.}~\bibnamefont
  {Bradlyn}}, \bibinfo {author} {\bibfnamefont {L.}~\bibnamefont {Elcoro}}, \
  and\ \bibinfo {author} {\bibfnamefont {e.~a.}\ \bibnamefont
  {Jennifer~Cano}},\ }\href {\doibase https://doi.org/10.1038/nature23268}
  {\bibfield  {journal} {\bibinfo  {journal} {Nature}\ }\textbf {\bibinfo
  {volume} {547}},\ \bibinfo {pages} {298} (\bibinfo {year}
  {2017})}\BibitemShut {NoStop}%
\bibitem [{\citenamefont {Schindler}\ \emph {et~al.}(2018)\citenamefont
  {Schindler}, \citenamefont {Cook}, \citenamefont {Vergniory}, \citenamefont
  {Wang}, \citenamefont {Parkin}, \citenamefont {Bernevig},\ and\ \citenamefont
  {Neupert}}]{schindler2018higher}%
  \BibitemOpen
  \bibfield  {author} {\bibinfo {author} {\bibfnamefont {F.}~\bibnamefont
  {Schindler}}, \bibinfo {author} {\bibfnamefont {A.~M.}\ \bibnamefont {Cook}},
  \bibinfo {author} {\bibfnamefont {M.~G.}\ \bibnamefont {Vergniory}}, \bibinfo
  {author} {\bibfnamefont {Z.}~\bibnamefont {Wang}}, \bibinfo {author}
  {\bibfnamefont {S.~S.~P.}\ \bibnamefont {Parkin}}, \bibinfo {author}
  {\bibfnamefont {B.~A.}\ \bibnamefont {Bernevig}}, \ and\ \bibinfo {author}
  {\bibfnamefont {T.}~\bibnamefont {Neupert}},\ }\href@noop {} {\bibfield
  {journal} {\bibinfo  {journal} {Science Advances}\ }\textbf {\bibinfo
  {volume} {4}},\ \bibinfo {pages} {eaat0346} (\bibinfo {year}
  {2018})}\BibitemShut {NoStop}%
\bibitem [{\citenamefont {Watanabe}\ and\ \citenamefont
  {Oshikawa}(2018)}]{watanabe2018}%
  \BibitemOpen
  \bibfield  {author} {\bibinfo {author} {\bibfnamefont {H.}~\bibnamefont
  {Watanabe}}\ and\ \bibinfo {author} {\bibfnamefont {M.}~\bibnamefont
  {Oshikawa}},\ }\href {\doibase 10.1103/physrevx.8.021065} {\bibfield
  {journal} {\bibinfo  {journal} {Physical Review X}\ }\textbf {\bibinfo
  {volume} {8}} (\bibinfo {year} {2018}),\
  10.1103/physrevx.8.021065}\BibitemShut {NoStop}%
\bibitem [{\citenamefont {van Miert}\ and\ \citenamefont
  {Ortix}(2018)}]{Miert2018dislocationCharge}%
  \BibitemOpen
  \bibfield  {author} {\bibinfo {author} {\bibfnamefont {G.}~\bibnamefont {van
  Miert}}\ and\ \bibinfo {author} {\bibfnamefont {C.}~\bibnamefont {Ortix}},\
  }\href {\doibase 10.1103/PhysRevB.97.201111} {\bibfield  {journal} {\bibinfo
  {journal} {Phys. Rev. B}\ }\textbf {\bibinfo {volume} {97}},\ \bibinfo
  {pages} {201111} (\bibinfo {year} {2018})}\BibitemShut {NoStop}%
\bibitem [{\citenamefont {Khalaf}\ \emph {et~al.}(2018)\citenamefont {Khalaf},
  \citenamefont {Po}, \citenamefont {Vishwanath},\ and\ \citenamefont
  {Watanabe}}]{khalaf2018symmetry}%
  \BibitemOpen
  \bibfield  {author} {\bibinfo {author} {\bibfnamefont {E.}~\bibnamefont
  {Khalaf}}, \bibinfo {author} {\bibfnamefont {H.~C.}\ \bibnamefont {Po}},
  \bibinfo {author} {\bibfnamefont {A.}~\bibnamefont {Vishwanath}}, \ and\
  \bibinfo {author} {\bibfnamefont {H.}~\bibnamefont {Watanabe}},\ }\href@noop
  {} {\bibfield  {journal} {\bibinfo  {journal} {Physical Review X}\ }\textbf
  {\bibinfo {volume} {8}},\ \bibinfo {pages} {031070} (\bibinfo {year}
  {2018})}\BibitemShut {NoStop}%
\bibitem [{\citenamefont {Thorngren}\ and\ \citenamefont
  {Else}(2018)}]{Thorngren2018}%
  \BibitemOpen
  \bibfield  {author} {\bibinfo {author} {\bibfnamefont {R.}~\bibnamefont
  {Thorngren}}\ and\ \bibinfo {author} {\bibfnamefont {D.~V.}\ \bibnamefont
  {Else}},\ }\href {\doibase 10.1103/PhysRevX.8.011040} {\bibfield  {journal}
  {\bibinfo  {journal} {Phys. Rev. X}\ }\textbf {\bibinfo {volume} {8}},\
  \bibinfo {pages} {011040} (\bibinfo {year} {2018})}\BibitemShut {NoStop}%
\bibitem [{\citenamefont {Tang}\ \emph {et~al.}(2019)\citenamefont {Tang},
  \citenamefont {Po}, \citenamefont {Vishwanath},\ and\ \citenamefont
  {Wan}}]{tang2019comprehensive}%
  \BibitemOpen
  \bibfield  {author} {\bibinfo {author} {\bibfnamefont {F.}~\bibnamefont
  {Tang}}, \bibinfo {author} {\bibfnamefont {H.~C.}\ \bibnamefont {Po}},
  \bibinfo {author} {\bibfnamefont {A.}~\bibnamefont {Vishwanath}}, \ and\
  \bibinfo {author} {\bibfnamefont {X.}~\bibnamefont {Wan}},\ }\href@noop {}
  {\bibfield  {journal} {\bibinfo  {journal} {Nature}\ }\textbf {\bibinfo
  {volume} {566}},\ \bibinfo {pages} {486} (\bibinfo {year}
  {2019})}\BibitemShut {NoStop}%
\bibitem [{\citenamefont {Liu}\ \emph {et~al.}(2019)\citenamefont {Liu},
  \citenamefont {Vishwanath},\ and\ \citenamefont {Khalaf}}]{Liu2019ShiftIns}%
  \BibitemOpen
  \bibfield  {author} {\bibinfo {author} {\bibfnamefont {S.}~\bibnamefont
  {Liu}}, \bibinfo {author} {\bibfnamefont {A.}~\bibnamefont {Vishwanath}}, \
  and\ \bibinfo {author} {\bibfnamefont {E.}~\bibnamefont {Khalaf}},\ }\href
  {\doibase 10.1103/PhysRevX.9.031003} {\bibfield  {journal} {\bibinfo
  {journal} {Phys. Rev. X}\ }\textbf {\bibinfo {volume} {9}},\ \bibinfo {pages}
  {031003} (\bibinfo {year} {2019})}\BibitemShut {NoStop}%
\bibitem [{\citenamefont {Song}\ \emph {et~al.}(2020)\citenamefont {Song},
  \citenamefont {Fang},\ and\ \citenamefont {Qi}}]{Song2020}%
  \BibitemOpen
  \bibfield  {author} {\bibinfo {author} {\bibfnamefont {Z.}~\bibnamefont
  {Song}}, \bibinfo {author} {\bibfnamefont {C.}~\bibnamefont {Fang}}, \ and\
  \bibinfo {author} {\bibfnamefont {Y.}~\bibnamefont {Qi}},\ }\href {\doibase
  10.1038/s41467-020-17685-5} {\bibfield  {journal} {\bibinfo  {journal}
  {Nature Communications}\ }\textbf {\bibinfo {volume} {11}} (\bibinfo {year}
  {2020}),\ 10.1038/s41467-020-17685-5}\BibitemShut {NoStop}%
\bibitem [{\citenamefont {Li}\ \emph {et~al.}(2020)\citenamefont {Li},
  \citenamefont {Zhu}, \citenamefont {Benalcazar},\ and\ \citenamefont
  {Hughes}}]{Li2020disc}%
  \BibitemOpen
  \bibfield  {author} {\bibinfo {author} {\bibfnamefont {T.}~\bibnamefont
  {Li}}, \bibinfo {author} {\bibfnamefont {P.}~\bibnamefont {Zhu}}, \bibinfo
  {author} {\bibfnamefont {W.~A.}\ \bibnamefont {Benalcazar}}, \ and\ \bibinfo
  {author} {\bibfnamefont {T.~L.}\ \bibnamefont {Hughes}},\ }\href {\doibase
  10.1103/PhysRevB.101.115115} {\bibfield  {journal} {\bibinfo  {journal}
  {Phys. Rev. B}\ }\textbf {\bibinfo {volume} {101}},\ \bibinfo {pages}
  {115115} (\bibinfo {year} {2020})}\BibitemShut {NoStop}%
\bibitem [{\citenamefont {Manjunath}\ and\ \citenamefont
  {Barkeshli}(2021)}]{manjunath2021cgt}%
  \BibitemOpen
  \bibfield  {author} {\bibinfo {author} {\bibfnamefont {N.}~\bibnamefont
  {Manjunath}}\ and\ \bibinfo {author} {\bibfnamefont {M.}~\bibnamefont
  {Barkeshli}},\ }\href {\doibase 10.1103/PhysRevResearch.3.013040} {\bibfield
  {journal} {\bibinfo  {journal} {Phys. Rev. Research}\ }\textbf {\bibinfo
  {volume} {3}},\ \bibinfo {pages} {013040} (\bibinfo {year}
  {2021})}\BibitemShut {NoStop}%
\bibitem [{\citenamefont {Manjunath}\ and\ \citenamefont
  {Barkeshli}(2020)}]{manjunath2020FQH}%
  \BibitemOpen
  \bibfield  {author} {\bibinfo {author} {\bibfnamefont {N.}~\bibnamefont
  {Manjunath}}\ and\ \bibinfo {author} {\bibfnamefont {M.}~\bibnamefont
  {Barkeshli}},\ }\href {\doibase 10.48550/arxiv.2012.11603} {\enquote
  {\bibinfo {title} {Classification of fractional quantum hall states with
  spatial symmetries},}\ } (\bibinfo {year} {2020}),\ \Eprint
  {http://arxiv.org/abs/2012.11603} {arXiv:2012.11603 [cond-mat.str-el]}
  \BibitemShut {NoStop}%
\bibitem [{\citenamefont {Cano}\ and\ \citenamefont
  {Bradlyn}(2021)}]{Cano_2021}%
  \BibitemOpen
  \bibfield  {author} {\bibinfo {author} {\bibfnamefont {J.}~\bibnamefont
  {Cano}}\ and\ \bibinfo {author} {\bibfnamefont {B.}~\bibnamefont {Bradlyn}},\
  }\href {\doibase 10.1146/annurev-conmatphys-041720-124134} {\bibfield
  {journal} {\bibinfo  {journal} {Annual Review of Condensed Matter Physics}\
  }\textbf {\bibinfo {volume} {12}},\ \bibinfo {pages} {225} (\bibinfo {year}
  {2021})}\BibitemShut {NoStop}%
\bibitem [{\citenamefont {Elcoro}\ \emph {et~al.}(2021)\citenamefont {Elcoro},
  \citenamefont {Wieder}, \citenamefont {Song}, \citenamefont {Xu},
  \citenamefont {Bradlyn},\ and\ \citenamefont {Bernevig}}]{Elcoro2021tqc}%
  \BibitemOpen
  \bibfield  {author} {\bibinfo {author} {\bibfnamefont {L.}~\bibnamefont
  {Elcoro}}, \bibinfo {author} {\bibfnamefont {B.}~\bibnamefont {Wieder}},
  \bibinfo {author} {\bibfnamefont {Z.}~\bibnamefont {Song}}, \bibinfo {author}
  {\bibfnamefont {Y.}~\bibnamefont {Xu}}, \bibinfo {author} {\bibfnamefont
  {B.}~\bibnamefont {Bradlyn}}, \ and\ \bibinfo {author} {\bibfnamefont
  {B.~A.}\ \bibnamefont {Bernevig}},\ }\href {\doibase
  https://doi.org/10.1038/s41467-021-26241-8} {\bibfield  {journal} {\bibinfo
  {journal} {Nature Communications}\ }\textbf {\bibinfo {volume} {12}}
  (\bibinfo {year} {2021}),\
  https://doi.org/10.1038/s41467-021-26241-8}\BibitemShut {NoStop}%
\bibitem [{\citenamefont {Manjunath}\ \emph
  {et~al.}(2023{\natexlab{a}})\citenamefont {Manjunath}, \citenamefont
  {Calvera},\ and\ \citenamefont {Barkeshli}}]{manjunath2022mzm}%
  \BibitemOpen
  \bibfield  {author} {\bibinfo {author} {\bibfnamefont {N.}~\bibnamefont
  {Manjunath}}, \bibinfo {author} {\bibfnamefont {V.}~\bibnamefont {Calvera}},
  \ and\ \bibinfo {author} {\bibfnamefont {M.}~\bibnamefont {Barkeshli}},\
  }\href {\doibase 10.1103/PhysRevB.107.165126} {\bibfield  {journal} {\bibinfo
   {journal} {Phys. Rev. B}\ }\textbf {\bibinfo {volume} {107}},\ \bibinfo
  {pages} {165126} (\bibinfo {year} {2023}{\natexlab{a}})}\BibitemShut
  {NoStop}%
\bibitem [{\citenamefont {Herzog-Arbeitman}\ \emph {et~al.}(2022)\citenamefont
  {Herzog-Arbeitman}, \citenamefont {Bernevig},\ and\ \citenamefont
  {Song}}]{herzogarbeitman2022interacting}%
  \BibitemOpen
  \bibfield  {author} {\bibinfo {author} {\bibfnamefont {J.}~\bibnamefont
  {Herzog-Arbeitman}}, \bibinfo {author} {\bibfnamefont {B.~A.}\ \bibnamefont
  {Bernevig}}, \ and\ \bibinfo {author} {\bibfnamefont {Z.-D.}\ \bibnamefont
  {Song}},\ }\href@noop {} {\enquote {\bibinfo {title} {Interacting topological
  quantum chemistry in 2d: Many-body real space invariants},}\ } (\bibinfo
  {year} {2022}),\ \Eprint {http://arxiv.org/abs/2212.00030} {arXiv:2212.00030
  [cond-mat.str-el]} \BibitemShut {NoStop}%
\bibitem [{\citenamefont {Zhang}\ \emph
  {et~al.}(2022{\natexlab{a}})\citenamefont {Zhang}, \citenamefont {Manjunath},
  \citenamefont {Nambiar},\ and\ \citenamefont
  {Barkeshli}}]{zhang2022fractional}%
  \BibitemOpen
  \bibfield  {author} {\bibinfo {author} {\bibfnamefont {Y.}~\bibnamefont
  {Zhang}}, \bibinfo {author} {\bibfnamefont {N.}~\bibnamefont {Manjunath}},
  \bibinfo {author} {\bibfnamefont {G.}~\bibnamefont {Nambiar}}, \ and\
  \bibinfo {author} {\bibfnamefont {M.}~\bibnamefont {Barkeshli}},\ }\href
  {\doibase 10.1103/PhysRevLett.129.275301} {\bibfield  {journal} {\bibinfo
  {journal} {Phys. Rev. Lett.}\ }\textbf {\bibinfo {volume} {129}},\ \bibinfo
  {pages} {275301} (\bibinfo {year} {2022}{\natexlab{a}})}\BibitemShut
  {NoStop}%
\bibitem [{\citenamefont {Zhang}\ \emph {et~al.}(2023)\citenamefont {Zhang},
  \citenamefont {Manjunath}, \citenamefont {Kobayashi},\ and\ \citenamefont
  {Barkeshli}}]{zhang2023complete}%
  \BibitemOpen
  \bibfield  {author} {\bibinfo {author} {\bibfnamefont {Y.}~\bibnamefont
  {Zhang}}, \bibinfo {author} {\bibfnamefont {N.}~\bibnamefont {Manjunath}},
  \bibinfo {author} {\bibfnamefont {R.}~\bibnamefont {Kobayashi}}, \ and\
  \bibinfo {author} {\bibfnamefont {M.}~\bibnamefont {Barkeshli}},\ }\href@noop
  {} {\bibfield  {journal} {\bibinfo  {journal} {Physical Review Letters}\
  }\textbf {\bibinfo {volume} {131}},\ \bibinfo {pages} {176501} (\bibinfo
  {year} {2023})}\BibitemShut {NoStop}%
\bibitem [{\citenamefont {Manjunath}\ \emph
  {et~al.}(2024{\natexlab{a}})\citenamefont {Manjunath}, \citenamefont
  {Calvera},\ and\ \citenamefont {Barkeshli}}]{manjunath2023classif}%
  \BibitemOpen
  \bibfield  {author} {\bibinfo {author} {\bibfnamefont {N.}~\bibnamefont
  {Manjunath}}, \bibinfo {author} {\bibfnamefont {V.}~\bibnamefont {Calvera}},
  \ and\ \bibinfo {author} {\bibfnamefont {M.}~\bibnamefont {Barkeshli}},\
  }\href {\doibase 10.1103/PhysRevB.109.035168} {\bibfield  {journal} {\bibinfo
   {journal} {Phys. Rev. B}\ }\textbf {\bibinfo {volume} {109}},\ \bibinfo
  {pages} {035168} (\bibinfo {year} {2024}{\natexlab{a}})}\BibitemShut
  {NoStop}%
\bibitem [{\citenamefont {Sachdev}(2023)}]{sachdev2023quantum}%
  \BibitemOpen
  \bibfield  {author} {\bibinfo {author} {\bibfnamefont {S.}~\bibnamefont
  {Sachdev}},\ }\href@noop {} {\emph {\bibinfo {title} {Quantum Phases of
  Matter}}}\ (\bibinfo  {publisher} {Cambridge University Press},\ \bibinfo
  {year} {2023})\BibitemShut {NoStop}%
\bibitem [{\citenamefont {Kobayashi}\ \emph
  {et~al.}(2024{\natexlab{a}})\citenamefont {Kobayashi}, \citenamefont {Zhang},
  \citenamefont {Wang},\ and\ \citenamefont {Barkeshli}}]{kobayashi2024}%
  \BibitemOpen
  \bibfield  {author} {\bibinfo {author} {\bibfnamefont {R.}~\bibnamefont
  {Kobayashi}}, \bibinfo {author} {\bibfnamefont {Y.}~\bibnamefont {Zhang}},
  \bibinfo {author} {\bibfnamefont {Y.-Q.}\ \bibnamefont {Wang}}, \ and\
  \bibinfo {author} {\bibfnamefont {M.}~\bibnamefont {Barkeshli}},\ }\href@noop
  {} {\enquote {\bibinfo {title} {(2+1)d topological phases with rt symmetry:
  many-body invariant, classification, and higher order edge modes},}\ }
  (\bibinfo {year} {2024}{\natexlab{a}}),\ \Eprint
  {http://arxiv.org/abs/2403.18887} {arXiv:2403.18887 [cond-mat.str-el]}
  \BibitemShut {NoStop}%
\bibitem [{\citenamefont {Zhang}\ \emph
  {et~al.}(2022{\natexlab{b}})\citenamefont {Zhang}, \citenamefont {Manjunath},
  \citenamefont {Nambiar},\ and\ \citenamefont {Barkeshli}}]{zhang2022pol}%
  \BibitemOpen
  \bibfield  {author} {\bibinfo {author} {\bibfnamefont {Y.}~\bibnamefont
  {Zhang}}, \bibinfo {author} {\bibfnamefont {N.}~\bibnamefont {Manjunath}},
  \bibinfo {author} {\bibfnamefont {G.}~\bibnamefont {Nambiar}}, \ and\
  \bibinfo {author} {\bibfnamefont {M.}~\bibnamefont {Barkeshli}},\ }\href
  {\doibase 10.48550/ARXIV.2211.09127} {\enquote {\bibinfo {title} {Quantized
  charge polarization as a many-body invariant in (2+1)d crystalline
  topological states and hofstadter butterflies},}\ } (\bibinfo {year}
  {2022}{\natexlab{b}}),\ \Eprint {http://arxiv.org/abs/2211.09127}
  {2211.09127} \BibitemShut {NoStop}%
\bibitem [{\citenamefont {Manjunath}\ \emph
  {et~al.}(2024{\natexlab{b}})\citenamefont {Manjunath}, \citenamefont
  {Calvera},\ and\ \citenamefont {Barkeshli}}]{manjunath2024Characterization}%
  \BibitemOpen
  \bibfield  {author} {\bibinfo {author} {\bibfnamefont {N.}~\bibnamefont
  {Manjunath}}, \bibinfo {author} {\bibfnamefont {V.}~\bibnamefont {Calvera}},
  \ and\ \bibinfo {author} {\bibfnamefont {M.}~\bibnamefont {Barkeshli}},\
  }\href {\doibase 10.1103/PhysRevB.109.035168} {\bibfield  {journal} {\bibinfo
   {journal} {Phys. Rev. B}\ }\textbf {\bibinfo {volume} {109}},\ \bibinfo
  {pages} {035168} (\bibinfo {year} {2024}{\natexlab{b}})}\BibitemShut
  {NoStop}%
\bibitem [{\citenamefont {Kobayashi}\ \emph
  {et~al.}(2024{\natexlab{b}})\citenamefont {Kobayashi}, \citenamefont {Zhang},
  \citenamefont {Manjunath},\ and\ \citenamefont
  {Barkeshli}}]{kobayashi2024crystalline}%
  \BibitemOpen
  \bibfield  {author} {\bibinfo {author} {\bibfnamefont {R.}~\bibnamefont
  {Kobayashi}}, \bibinfo {author} {\bibfnamefont {Y.}~\bibnamefont {Zhang}},
  \bibinfo {author} {\bibfnamefont {N.}~\bibnamefont {Manjunath}}, \ and\
  \bibinfo {author} {\bibfnamefont {M.}~\bibnamefont {Barkeshli}},\ }\href
  {https://arxiv.org/abs/2405.17431} {\enquote {\bibinfo {title} {Crystalline
  invariants of fractional chern insulators},}\ } (\bibinfo {year}
  {2024}{\natexlab{b}}),\ \Eprint {http://arxiv.org/abs/2405.17431}
  {arXiv:2405.17431 [cond-mat.str-el]} \BibitemShut {NoStop}%
\bibitem [{\citenamefont {Song}\ \emph {et~al.}(2021)\citenamefont {Song},
  \citenamefont {He}, \citenamefont {Vishwanath},\ and\ \citenamefont
  {Wang}}]{Song2021polarization}%
  \BibitemOpen
  \bibfield  {author} {\bibinfo {author} {\bibfnamefont {X.-Y.}\ \bibnamefont
  {Song}}, \bibinfo {author} {\bibfnamefont {Y.-C.}\ \bibnamefont {He}},
  \bibinfo {author} {\bibfnamefont {A.}~\bibnamefont {Vishwanath}}, \ and\
  \bibinfo {author} {\bibfnamefont {C.}~\bibnamefont {Wang}},\ }\href {\doibase
  10.1103/PhysRevResearch.3.023011} {\bibfield  {journal} {\bibinfo  {journal}
  {Phys. Rev. Research}\ }\textbf {\bibinfo {volume} {3}},\ \bibinfo {pages}
  {023011} (\bibinfo {year} {2021})}\BibitemShut {NoStop}%
\bibitem [{\citenamefont {Coh}\ and\ \citenamefont
  {Vanderbilt}(2009)}]{coh2009}%
  \BibitemOpen
  \bibfield  {author} {\bibinfo {author} {\bibfnamefont {S.}~\bibnamefont
  {Coh}}\ and\ \bibinfo {author} {\bibfnamefont {D.}~\bibnamefont
  {Vanderbilt}},\ }\href {\doibase 10.1103/PhysRevLett.102.107603} {\bibfield
  {journal} {\bibinfo  {journal} {Phys. Rev. Lett.}\ }\textbf {\bibinfo
  {volume} {102}},\ \bibinfo {pages} {107603} (\bibinfo {year}
  {2009})}\BibitemShut {NoStop}%
\bibitem [{\citenamefont {Fang}\ \emph {et~al.}(2012)\citenamefont {Fang},
  \citenamefont {Gilbert},\ and\ \citenamefont {Bernevig}}]{Fang2012PGS}%
  \BibitemOpen
  \bibfield  {author} {\bibinfo {author} {\bibfnamefont {C.}~\bibnamefont
  {Fang}}, \bibinfo {author} {\bibfnamefont {M.~J.}\ \bibnamefont {Gilbert}}, \
  and\ \bibinfo {author} {\bibfnamefont {B.~A.}\ \bibnamefont {Bernevig}},\
  }\href {\doibase 10.1103/PhysRevB.86.115112} {\bibfield  {journal} {\bibinfo
  {journal} {Phys. Rev. B}\ }\textbf {\bibinfo {volume} {86}},\ \bibinfo
  {pages} {115112} (\bibinfo {year} {2012})}\BibitemShut {NoStop}%
\bibitem [{\citenamefont {Benalcazar}\ \emph {et~al.}(2019)\citenamefont
  {Benalcazar}, \citenamefont {Li},\ and\ \citenamefont
  {Hughes}}]{Benalcazar2019HOTI}%
  \BibitemOpen
  \bibfield  {author} {\bibinfo {author} {\bibfnamefont {W.~A.}\ \bibnamefont
  {Benalcazar}}, \bibinfo {author} {\bibfnamefont {T.}~\bibnamefont {Li}}, \
  and\ \bibinfo {author} {\bibfnamefont {T.~L.}\ \bibnamefont {Hughes}},\
  }\href {\doibase 10.1103/PhysRevB.99.245151} {\bibfield  {journal} {\bibinfo
  {journal} {Phys. Rev. B}\ }\textbf {\bibinfo {volume} {99}},\ \bibinfo
  {pages} {245151} (\bibinfo {year} {2019})}\BibitemShut {NoStop}%
\bibitem [{\citenamefont {Manjunath}\ \emph
  {et~al.}(2023{\natexlab{b}})\citenamefont {Manjunath}, \citenamefont {Prem},\
  and\ \citenamefont {Lu}}]{naren2023rotational}%
  \BibitemOpen
  \bibfield  {author} {\bibinfo {author} {\bibfnamefont {N.}~\bibnamefont
  {Manjunath}}, \bibinfo {author} {\bibfnamefont {A.}~\bibnamefont {Prem}}, \
  and\ \bibinfo {author} {\bibfnamefont {Y.-M.}\ \bibnamefont {Lu}},\ }\href
  {\doibase 10.1103/PhysRevB.107.195130} {\bibfield  {journal} {\bibinfo
  {journal} {Phys. Rev. B}\ }\textbf {\bibinfo {volume} {107}},\ \bibinfo
  {pages} {195130} (\bibinfo {year} {2023}{\natexlab{b}})}\BibitemShut
  {NoStop}%
\bibitem [{\citenamefont {Rao}\ and\ \citenamefont
  {Bradlyn}(2023)}]{rao2023effective}%
  \BibitemOpen
  \bibfield  {author} {\bibinfo {author} {\bibfnamefont {P.}~\bibnamefont
  {Rao}}\ and\ \bibinfo {author} {\bibfnamefont {B.}~\bibnamefont {Bradlyn}},\
  }\href {\doibase 10.1103/PhysRevB.107.195153} {\bibfield  {journal} {\bibinfo
   {journal} {Phys. Rev. B}\ }\textbf {\bibinfo {volume} {107}},\ \bibinfo
  {pages} {195153} (\bibinfo {year} {2023})}\BibitemShut {NoStop}%
\bibitem [{\citenamefont {May-Mann}\ and\ \citenamefont
  {Hughes}(2022)}]{maymann2022prb}%
  \BibitemOpen
  \bibfield  {author} {\bibinfo {author} {\bibfnamefont {J.}~\bibnamefont
  {May-Mann}}\ and\ \bibinfo {author} {\bibfnamefont {T.~L.}\ \bibnamefont
  {Hughes}},\ }\href {\doibase 10.1103/PhysRevB.106.L241113} {\bibfield
  {journal} {\bibinfo  {journal} {Phys. Rev. B}\ }\textbf {\bibinfo {volume}
  {106}},\ \bibinfo {pages} {L241113} (\bibinfo {year} {2022})}\BibitemShut
  {NoStop}%
\bibitem [{\citenamefont {Benalcazar}\ \emph {et~al.}(2017)\citenamefont
  {Benalcazar}, \citenamefont {Bernevig},\ and\ \citenamefont
  {Hughes}}]{wladimir2017quantized}%
  \BibitemOpen
  \bibfield  {author} {\bibinfo {author} {\bibfnamefont {W.~A.}\ \bibnamefont
  {Benalcazar}}, \bibinfo {author} {\bibfnamefont {B.~A.}\ \bibnamefont
  {Bernevig}}, \ and\ \bibinfo {author} {\bibfnamefont {T.~L.}\ \bibnamefont
  {Hughes}},\ }\href {\doibase 10.1126/science.aah6442} {\bibfield  {journal}
  {\bibinfo  {journal} {Science}\ }\textbf {\bibinfo {volume} {357}},\ \bibinfo
  {pages} {61} (\bibinfo {year} {2017})},\ \Eprint
  {http://arxiv.org/abs/https://www.science.org/doi/pdf/10.1126/science.aah6442}
  {https://www.science.org/doi/pdf/10.1126/science.aah6442} \BibitemShut
  {NoStop}%
\bibitem [{\citenamefont {Roy}\ and\ \citenamefont {Juri\ifmmode \check{c}\else
  \v{c}\fi{}i\ifmmode~\acute{c}\else \'{c}\fi{}}(2021)}]{roy2020dislocation}%
  \BibitemOpen
  \bibfield  {author} {\bibinfo {author} {\bibfnamefont {B.}~\bibnamefont
  {Roy}}\ and\ \bibinfo {author} {\bibfnamefont {V.}~\bibnamefont {Juri\ifmmode
  \check{c}\else \v{c}\fi{}i\ifmmode~\acute{c}\else \'{c}\fi{}}},\ }\href
  {\doibase 10.1103/PhysRevResearch.3.033107} {\bibfield  {journal} {\bibinfo
  {journal} {Phys. Rev. Res.}\ }\textbf {\bibinfo {volume} {3}},\ \bibinfo
  {pages} {033107} (\bibinfo {year} {2021})}\BibitemShut {NoStop}%
\bibitem [{\citenamefont {Hirsbrunner}\ \emph {et~al.}(2023)\citenamefont
  {Hirsbrunner}, \citenamefont {Gray},\ and\ \citenamefont
  {Hughes}}]{hirsbrunner2023crystalline}%
  \BibitemOpen
  \bibfield  {author} {\bibinfo {author} {\bibfnamefont {M.~R.}\ \bibnamefont
  {Hirsbrunner}}, \bibinfo {author} {\bibfnamefont {A.~D.}\ \bibnamefont
  {Gray}}, \ and\ \bibinfo {author} {\bibfnamefont {T.~L.}\ \bibnamefont
  {Hughes}},\ }\href@noop {} {\enquote {\bibinfo {title}
  {Crystalline-electromagnetic responses of higher order topological
  semimetals},}\ } (\bibinfo {year} {2023}),\ \Eprint
  {http://arxiv.org/abs/2308.05796} {arXiv:2308.05796 [cond-mat.mes-hall]}
  \BibitemShut {NoStop}%
\bibitem [{\citenamefont {Khalaf}(2018)}]{khalaf2018higher}%
  \BibitemOpen
  \bibfield  {author} {\bibinfo {author} {\bibfnamefont {E.}~\bibnamefont
  {Khalaf}},\ }\href@noop {} {\bibfield  {journal} {\bibinfo  {journal}
  {Physical Review B}\ }\textbf {\bibinfo {volume} {97}},\ \bibinfo {pages}
  {205136} (\bibinfo {year} {2018})}\BibitemShut {NoStop}%
\bibitem [{\citenamefont {Jahin}\ \emph {et~al.}(2024)\citenamefont {Jahin},
  \citenamefont {Lu},\ and\ \citenamefont {Wang}}]{Jahin2024HOTI}%
  \BibitemOpen
  \bibfield  {author} {\bibinfo {author} {\bibfnamefont {A.}~\bibnamefont
  {Jahin}}, \bibinfo {author} {\bibfnamefont {Y.-M.}\ \bibnamefont {Lu}}, \
  and\ \bibinfo {author} {\bibfnamefont {Y.}~\bibnamefont {Wang}},\ }\href
  {\doibase 10.1103/PhysRevB.109.205123} {\bibfield  {journal} {\bibinfo
  {journal} {Phys. Rev. B}\ }\textbf {\bibinfo {volume} {109}},\ \bibinfo
  {pages} {205123} (\bibinfo {year} {2024})}\BibitemShut {NoStop}%
\bibitem [{\citenamefont {Barkeshli}\ \emph {et~al.}(2022)\citenamefont
  {Barkeshli}, \citenamefont {Chen}, \citenamefont {Hsin},\ and\ \citenamefont
  {Manjunath}}]{barkeshli2021invertible}%
  \BibitemOpen
  \bibfield  {author} {\bibinfo {author} {\bibfnamefont {M.}~\bibnamefont
  {Barkeshli}}, \bibinfo {author} {\bibfnamefont {Y.-A.}\ \bibnamefont {Chen}},
  \bibinfo {author} {\bibfnamefont {P.-S.}\ \bibnamefont {Hsin}}, \ and\
  \bibinfo {author} {\bibfnamefont {N.}~\bibnamefont {Manjunath}},\ }\href
  {\doibase 10.1103/PhysRevB.105.235143} {\bibfield  {journal} {\bibinfo
  {journal} {Phys. Rev. B}\ }\textbf {\bibinfo {volume} {105}},\ \bibinfo
  {pages} {235143} (\bibinfo {year} {2022})}\BibitemShut {NoStop}%
\bibitem [{\citenamefont {Dijkgraaf}\ and\ \citenamefont
  {Witten}(1990)}]{dijkgraaf1990}%
  \BibitemOpen
  \bibfield  {author} {\bibinfo {author} {\bibfnamefont {R.}~\bibnamefont
  {Dijkgraaf}}\ and\ \bibinfo {author} {\bibfnamefont {E.}~\bibnamefont
  {Witten}},\ }\href@noop {} {\bibfield  {journal} {\bibinfo  {journal} {Comm.
  Math. Phys.}\ }\textbf {\bibinfo {volume} {129}},\ \bibinfo {pages} {393}
  (\bibinfo {year} {1990})}\BibitemShut {NoStop}%
\bibitem [{\citenamefont {Chen}\ \emph {et~al.}(2013)\citenamefont {Chen},
  \citenamefont {Gu}, \citenamefont {Liu},\ and\ \citenamefont
  {Wen}}]{Chen2013}%
  \BibitemOpen
  \bibfield  {author} {\bibinfo {author} {\bibfnamefont {X.}~\bibnamefont
  {Chen}}, \bibinfo {author} {\bibfnamefont {Z.-C.}\ \bibnamefont {Gu}},
  \bibinfo {author} {\bibfnamefont {Z.-X.}\ \bibnamefont {Liu}}, \ and\
  \bibinfo {author} {\bibfnamefont {X.-G.}\ \bibnamefont {Wen}},\ }\href
  {\doibase 10.1103/PhysRevB.87.155114} {\bibfield  {journal} {\bibinfo
  {journal} {Phys. Rev. B}\ }\textbf {\bibinfo {volume} {87}},\ \bibinfo
  {pages} {155114} (\bibinfo {year} {2013})}\BibitemShut {NoStop}%
\bibitem [{\citenamefont {Wen}\ and\ \citenamefont {Zee}(1992)}]{Wen1992shift}%
  \BibitemOpen
  \bibfield  {author} {\bibinfo {author} {\bibfnamefont {X.~G.}\ \bibnamefont
  {Wen}}\ and\ \bibinfo {author} {\bibfnamefont {A.}~\bibnamefont {Zee}},\
  }\href {\doibase 10.1103/PhysRevLett.69.953} {\bibfield  {journal} {\bibinfo
  {journal} {Phys. Rev. Lett.}\ }\textbf {\bibinfo {volume} {69}},\ \bibinfo
  {pages} {953} (\bibinfo {year} {1992})}\BibitemShut {NoStop}%
\bibitem [{\citenamefont {Abanov}\ and\ \citenamefont
  {Gromov}(2014)}]{Gromov2014}%
  \BibitemOpen
  \bibfield  {author} {\bibinfo {author} {\bibfnamefont {A.~G.}\ \bibnamefont
  {Abanov}}\ and\ \bibinfo {author} {\bibfnamefont {A.}~\bibnamefont
  {Gromov}},\ }\href {\doibase 10.1103/PhysRevB.90.014435} {\bibfield
  {journal} {\bibinfo  {journal} {Phys. Rev. B}\ }\textbf {\bibinfo {volume}
  {90}},\ \bibinfo {pages} {014435} (\bibinfo {year} {2014})}\BibitemShut
  {NoStop}%
\bibitem [{\citenamefont {Gromov}\ \emph {et~al.}(2015)\citenamefont {Gromov},
  \citenamefont {Cho}, \citenamefont {You}, \citenamefont {Abanov},\ and\
  \citenamefont {Fradkin}}]{Gromov2015}%
  \BibitemOpen
  \bibfield  {author} {\bibinfo {author} {\bibfnamefont {A.}~\bibnamefont
  {Gromov}}, \bibinfo {author} {\bibfnamefont {G.~Y.}\ \bibnamefont {Cho}},
  \bibinfo {author} {\bibfnamefont {Y.}~\bibnamefont {You}}, \bibinfo {author}
  {\bibfnamefont {A.~G.}\ \bibnamefont {Abanov}}, \ and\ \bibinfo {author}
  {\bibfnamefont {E.}~\bibnamefont {Fradkin}},\ }\href {\doibase
  10.1103/PhysRevLett.114.016805} {\bibfield  {journal} {\bibinfo  {journal}
  {Phys. Rev. Lett.}\ }\textbf {\bibinfo {volume} {114}},\ \bibinfo {pages}
  {016805} (\bibinfo {year} {2015})}\BibitemShut {NoStop}%
\bibitem [{\citenamefont {Gromov}\ \emph {et~al.}(2016)\citenamefont {Gromov},
  \citenamefont {Jensen},\ and\ \citenamefont {Abanov}}]{gromov2016boundary}%
  \BibitemOpen
  \bibfield  {author} {\bibinfo {author} {\bibfnamefont {A.}~\bibnamefont
  {Gromov}}, \bibinfo {author} {\bibfnamefont {K.}~\bibnamefont {Jensen}}, \
  and\ \bibinfo {author} {\bibfnamefont {A.~G.}\ \bibnamefont {Abanov}},\
  }\href {\doibase 10.1103/PhysRevLett.116.126802} {\bibfield  {journal}
  {\bibinfo  {journal} {Phys. Rev. Lett.}\ }\textbf {\bibinfo {volume} {116}},\
  \bibinfo {pages} {126802} (\bibinfo {year} {2016})}\BibitemShut {NoStop}%
\end{thebibliography}%
\clearpage

\end{document}